\documentclass[aps,preprint,floatfix,nofootinbib,showpacs,a4paper]{revtex4-1}
\usepackage{graphicx}
\usepackage[dvipsnames]{xcolor}

\usepackage{cancel}
\usepackage[normalem]{ulem}
\usepackage{multirow}
\usepackage{amsmath,amssymb}
\usepackage{hyperref}
\usepackage{soul}

\newcommand{\gev}{\ensuremath{\,\mathrm{GeV}}}
\newcommand{\tev}{\ensuremath{\,\mathrm{TeV}}}

\newcommand{\mET}{\cancel{\it{E}}_{T}}

\def\ga   {\gamma}

\def\th   {\theta}
\def\la   {\lambda}

\def\nn{\nonumber}
\def\lee { \left( }
\def\rii { \right) }
\def\lan   {\langle}
\def\ran   {\rangle}

\def\De {\Delta}

\def\to {\rightarrow}

\begin{document}

{\small
\begin{flushright}
DO-TH 17/18,
NCTS-PH/1718\\
CP3-Origins-2017-031 DNRF90\\
\end{flushright} }

\medskip

\title{
Signals of New Gauge Bosons in \\
Gauged Two Higgs Doublet Model}

\renewcommand{\thefootnote}{\fnsymbol{footnote}}

\author{
Wei-Chih Huang$^{1,2}$, Hiroyuki Ishida$^{3}$, \\
Chih-Ting Lu$^{4}$, Yue-Lin Sming Tsai$^{3}$ and Tzu-Chiang Yuan$^{3,5}$}
\affiliation{
$^1$ Fakult\"at f\"ur Physik, Technische Universit\"at Dortmund,
44221 Dortmund, Germany \\
$^2$ CP$^{\,3}$-Origins, University of Southern Denmark, Campusvej 55, DK-5230 Odense M, Denmark \\
$^3$ Physics Division, National Center for Theoretical Sciences, Hsinchu 300, Taiwan \\
$^4$ Department of Physics, National Tsing Hua University,
Hsinchu 300, Taiwan \\
$^5$ Institute of Physics, Academia Sinica, Nangang, Taipei 11529, Taiwan 
}
\date{\today}

\begin{abstract}
Recently a gauged two Higgs doublet model, in which the two Higgs doublets 
are embedded into the fundamental representation of an extra local $SU(2)_H$ group, is constructed. 
Both the new gauge bosons $Z^\prime$ and $W^{\prime (p,m)}$ are electrically neutral.
While $Z^\prime$ can be singly produced at colliders, $W^{\prime (p,m)}$, which is heavier, must be pair produced.
We explore the constraints of $Z^\prime$ using the 
current Drell-Yan type data from the 
Large Hadron Collider. Anticipating optimistically that $Z^\prime$ can be 
discovered via the clean Drell-Yan type signals at high luminosity upgrade 
of the collider, we explore the detectability of extra heavy 
fermions in the model via the two leptons/jets plus missing transverse energy signals from the 
exotic decay modes of $Z^\prime$.
For the $W^{\prime (p,m)}$ pair production in a future 100 TeV proton-proton collider, 
we demonstrate certain kinematical distributions for the two/four leptons 
plus missing energy signals have distinguishable features from the Standard Model background.  
In addition, comparisons of these kinematical distributions between the gauged two Higgs doublet 
model and the littlest Higgs model with T-parity, the latter of which can give rise to the same signals with competitive if not larger cross sections, are also presented. 

\end{abstract}
\maketitle

\section{Introduction}

With the discovery of the 125 GeV scalar boson at the Large Hadron Collider (LHC), the Standard Model (SM) with just one Higgs doublet has now been generally accepted as the standard theory or framework to describe the fundamental strong and electroweak interactions for three generations of elementary particles of quarks and leptons.
An extended Higgs sector, however, is often used to address various theoretical puzzles like neutrino masses, 
dark matter (DM), matter-antimatter asymmetry, hierarchy problem {\it etc.} which remain unexplained in this standard framework. 
Perhaps the general two Higgs doublet model (2HDM), 
in particular in the context of supersymmetric theories, 
is the most studied in the literature. Due to its diverse variations,
2HDM has been used as a prototype to address aforementioned theoretical issues. 
For reviews of 2HDM and its supersymmetric version, 
see for example~\cite{Branco:2011iw,Djouadi:2005gi,Gunion:1989we}.
One of the interesting 2HDM variants is the inert Higgs doublet model~\cite{Deshpande:1977rw,Ma:2006km,Barbieri:2006dq,LopezHonorez:2006gr}, 
in which the neutral component of the
second Higgs doublet can be a dark matter candidate due to a discrete $Z_2$ symmetry imposed 
on the scalar potential. Detailed analysis for the inert Higgs doublet model can be found for example in~\cite{Arhrib:2013ela,Ilnicka:2015jba,Belyaev:2016lok,Eiteneuer:2017hoh}. Origin of multiple inert Higgs doublets in the context of grand unification 
has been addressed in~\cite{Kephart:2015oaa}.

In a recent work~\cite{Huang:2015wts}, we have proposed a novel model, dubbed Gauged Two Higgs Doublet Model~(G2HDM), where the two Higgs doublets $H_1$ and $H_2$ in 2HDM are embedded 
into a doublet $H=(H_1,H_2)^T$ of a new non-abelian $SU(2)_H$ gauge group.
The SM $SU(2)_L$ right-handed singlet fermions are paired up with new fermions to form $SU(2)_H$ doublets, 
whereas $SU(2)_L$ left-handed doublet fermions are singlets under the $SU(2)_H$. 
Four additional chiral (left-handed) fermions for each generation, 
all singlets under both $SU(2)_L$ and $SU(2)_H$,   
are also introduced to render the model free of gauge anomalies.
In this model, an inert Higgs doublet can be naturally realized without imposing the ad-hoc $Z_2$ symmetry mentioned above
to accommodate a DM candidate. Flavor changing neutral currents  
are also absent naturally at tree level. We note that it is widely believed that global symmetry (whether it is discrete or continuous)
may be strongly violated by gravitational effects~\cite{Krauss:1988zc,Kallosh:1995hi}. Thus from an effective field theory point of view it is desirable to embed discrete symmetries into local gauge symmetries below the Planck scale~\cite{Krauss:1988zc}. 
Indeed, besides $Z_2$,  non-abelian discrete flavor groups like
$\{ A_4, S_4, A_5 \}$, $\{Q_6,  T^\prime, O^\prime, I^\prime \}$, and $\{ T_7, \Delta(27), PSL(2,7) \}$ 
can be minimally embedded into $SO(3)$, $SU(2)$, and $SU(3)$ respectively~\cite{Rachlin:2017rvm}.

In G2HDM, a distinctive feature is all the $SU(2)_H$ gauge bosons $Z^\prime$ 
and $W^{\prime\, (p,m)}$ are electrically neutral which is not the case for 
the Left-Right symmetry model (LRSM)~\cite{LRSM},  
the littlest Higgs model with T-parity (LHT)~\cite{Cheng:2003ju}, and 
the original Twin Higgs model (THM)~\cite{Chacko:2005pe}. 
Naturally one might ask how do we distinguish the G2HDM $Z'$ and $W^{\prime\, (p,m)}$ 
from other gauge boson impostors which also arise from the non-abelian group $SU(2)$?
In terms of collider searches, some of the new gauge bosons from the aforementioned models
can never be singly produced due to the gauge symmetry involved.
For instance, the $W^{\prime\, (p,m)}$ in G2HDM and $W_H^\pm$ in LHT may 
{\it not} be singly produced at the LHC. 
Generally speaking, the $Z'$ and $W^{\prime\, (p,m)}$ as well as their impostors are short-lived 
and it is not possible to identify the new gauge bosons using the tracking or displaced vertex techniques designed 
for long-lived particles.
Hence, in additional to their production cross sections, detailed kinematics distributions have 
to be involved for making differentiation.

If new gauge bosons can be singly produced, 
they will be stringently constrained by exotic searches from 
the LHC~\cite{Khachatryan:2014fba,ATLAS:2016lvi} due to large~(resonant) production cross sections. 
The latest LHC 13 TeV $Z'$ resonance searches based on the channels of dilepton~\cite{ATLAS:2016cyf, CMS:2016abv}, 
dijet~\cite{ATLAS:2016lvi, CMS:2016wpz}, $b$-quark pair~\cite{ATLAS:2016gvq}, 
$t$-quark pair~\cite{TheATLAScollaboration:2016wfb}, and 
other bosonic final states~\cite{ATLAS:2016kxc, ATLAS:2016cwq, ATLAS:2016yqq} 
have recently been released. 
Among these searches, the cleanest dilepton channels yield the most strong constraint on the 
$Z^\prime$ coupling to SM fermions 
in light of small background.
Moreover, from the total electric charge of the decay products it is straightforward 
to tell singly produced charged bosons from neutral ones.

The current bound from the LHC dilepton searches~\cite{ATLAS:2016cyf, CMS:2016abv} 
for $Z_R$ in the minimal LRSM, assuming  
$m_{Z_R} = 1.7 \, m_{W_{R}}$ and $g_L = g_R$, is $m_{Z^\prime_R}> 3.2$ TeV~\cite{Lindner:2016lpp}. 
Similarly for $Z^\prime$ and $W^\prime$ in the Left-Right THM (LRTH)~\cite{Goh:2006wj} with 
$m_{W^\prime} = m_{Z^\prime} \sqrt{\cos 2\theta_w}/\cos\theta_w$ and $g_L = g_R$, one obtains the limit 
$m_{Z^\prime} > 3.36 \, \tev$.
In some variants of the THM~(see e.g. Ref.~\cite{Craig:2015pha}), where
the $SU(2)$ symmetry is doubled and the $U(1)$ symmetry of 
the twin sector becomes a global symmetry, 
the exotic $W$ boson can only be doubly produced. 
In this case, the exotic $W$ might behave similarly like $W^{\prime\, (p,m)}$ in G2HDM in terms of collider signatures.
To distinguish them, one will need more information such as the new scalar and fermion mass spectra which are quite different in the two models.

Suppose a new $Z'$ is observed via resonance searches at the LHC. We would like to know if this new 
$Z'$ belongs to a new abelian $U(1)$ or a member of a new $SU(2)$.
In order to confirm the existence of G2HDM, the next step is to discover the neutral $W^{\prime \, (p,m)}$.
In G2HDM, unlike $Z'$, the $W^{\prime \, (p,m)}$ do not couple bilinearly to the SM quarks. Thus 
$Z'$ can be singly produced via quark-antiquark annihilations 
while $W^{\prime \, (p,m)}$ must be pair produced via exchange of new heavy fermions or $Z^\prime$.
For $W^{\prime \, (p,m)}$ produced in pairs, more information like detailed kinematical distributions 
other than the production cross sections have to be involved so as to make distinguishable signatures
from say the $W^\pm_H$ pair in LHT.

Besides the new gauge bosons,  in order to give masses to all fermions the scalar sector is also enlarged beyond the two Higgs doublets with one  extra doublet $\Phi_H$ and one extra triplet $\Delta_H$ of $SU(2)_H$, 
both of which are singlets under $SU(2)_L$. 
The particle content of the G2HDM~\cite{Huang:2015wts} is summarized in Table~\ref{tab:quantumnos} together
with their quantum numbers. 

In this work, we will focus on two benchmark mass spectra (Spectrum-A and Spectrum-B) of the G2HDM 
for our collider studies.  
The Spectrum-A contains heavy 
and decoupled new quarks while the Spectrum-B comprises relatively light new quarks.
For all scenarios, new leptons are assumed to be lighter than the additional gauge bosons of interest.
Due to the fact $Z^\prime$ couples to SM quarks and can be singly produced at the LHC, we first update the bounds on the $SU(2)_H$ gauge 
coupling $g_H$ as a function of the $Z^\prime$ mass  $m_{Z^\prime}$ by using the newly released results of the dilepton and dijet searches from the LHC.
Next, $Z^\prime$ exotic decays into new heavy fermions followed by decays into SM fermions are investigated at the 14 TeV  High Luminosity LHC (HL-LHC)
and bounds from LHC searches on supersymmetric particles can be applied with simplified assumptions. 
Then, for the neutral $W^{\prime \,(p,m)}$ in G2HDM   
we propose searching for two channels: two leptons and four leptons with missing transverse energy.
We shall demonstrate that the pair production of $W^{\prime \,(p,m)}$ can feature quite 
distinctive kinematical distributions from the $W^\prime_H$ pair in LHT, which will be chosen as
a representative model for comparisons since $W^\prime$ can only be pair produced in both models.

This rest of this paper is laid out as follows. In Sec.~\ref{sec:G2HDM}, 
we briefly review the G2HDM and spell out the relevant gauge
interactions for collider searches of interest. 
In Sec.~\ref{sec:methodology}, we discuss the methodology employed in the collider simulations.
In Sec.~\ref{sec:Zprime}, we revisit $Z'$ direct search limits from  
the latest $13\tev$ LHC data as well as exploring some of its exotic decay channels at the HL-LHC.
In Sec.~\ref{sec:Wprime}, signatures for $W^\prime$ at a future 100 TeV proton-proton 
collider are scrutinized in detail and compared with those from LHT.
We summarize our findings and conclude in Sec.~\ref{sec:summary}.  
For convenience, we also present the scalar potential of G2HDM and the associated scalar mass spectra in two appendixes.
More details of the scalar sector of G2HDM can be found in~\cite{Arhrib:2018sbz}.

\begin{table}[htp!]
\begin{tabular}{|c|c|c|c|c||c|}
\hline
Matter Fields & $SU(3)_C$ & $SU(2)_L$ & $SU(2)_H$ & $U(1)_Y$ & $U(1)_X$ \\
\hline \hline
$H=\left( H_1 \,,\, H_2 \right)^T$ & 1 & 2 & 2 & 1/2 & $1$ \\
$\Phi_H=\left( \Phi_1 \,,\, \Phi_2 \right)^T$ & 1 & 1 & 2 & 0 & $1$ \\
$\Delta_H=\left( \begin{array}{cc} \Delta_3/2 & \Delta_p/\sqrt{2}  \\ \Delta_m/\sqrt{2} & - \Delta_3/2 \end{array} \right)$ & 1 & 1 & 3 & 0 & 0 \\
\hline\hline
$Q_L=\left( u_L \,,\, d_L \right)^T$ & 3 & 2 & 1 & 1/6 & 0\\
$U_R=\left( u_R \,,\, u^H_R \right)^T$ & 3 & 1 & 2 & 2/3 & $1$ \\
$D_R=\left( d^H_R \,,\, d_R \right)^T$ & 3 & 1 & 2 & $-1/3$ & $-1$ \\
\hline
$u^H_L$ & 3 & 1 & 1 & 2/3 & 0 \\
$d^H_L$ & 3 & 1 & 1 & $-1/3$ & 0 \\
\hline
$L_L=\left( \nu_L \,,\, e_L \right)^T$ & 1 & 2 & 1 & $-1/2$ & 0 \\
$N_R=\left( \nu_R \,,\, \nu^H_R \right)^T$ & 1 & 1 & 2 & 0 & $1$ \\
$E_R=\left( e^H_R \,,\, e_R \right)^T$ & 1 & 1 & 2 &  $-1$  &  $-1$ \\
\hline
$\nu^H_L$ & 1 & 1 & 1 & 0 & 0 \\
$e^H_L$ & 1 & 1 & 1 & $-1$ & 0 \\
\hline
\end{tabular}
\caption{Matter field contents and their quantum number assignments in G2HDM. 
}
\label{tab:quantumnos}
\end{table}

\section{G2HDM gauge interactions}
\label{sec:G2HDM}

In this section, we give a brief review on G2HDM, focusing on gauge interactions that are relevant to our study of collider searches.
The particle contents summarized in Table~\ref{tab:quantumnos} have the minimal set of 
new heavy chiral fermions required for anomaly cancellation and new scalars for facilitating spontaneous 
electroweak symmetry breaking, as proposed in~\cite{Huang:2015wts}.   

As mentioned earlier, the two $SU(2)_L$ Higgs doublets $H_1$ and $H_2$ 
are embedded into a doublet $H$ under a non-abelian $SU(2)_H$ gauge group. 
$H$ is also charged under an additional gauged abelian group $U(1)_X$.
To provide masses to the additional gauge bosons, we introduce an $SU(2)_H$ scalar triplet $\Delta_H$ 
and doublet $\Phi_H$ (both are singlets under the SM gauge group).  
The vacuum expectation value~(vev) of the triplet $\Delta_H$ not only breaks $SU(2)_H$ spontaneously, 
but it also triggers the electroweak symmetry breaking by inducing a vev to the first $SU(2)_L$ doublet $H_1$, which is identified as the SM Higgs doublet.
In contrast, the second Higgs doublet $H_2$ does not obtain a vev and
its neutral component could be the DM candidate. As shown in~\cite{Huang:2015wts}, 
DM stability is protected by the $SU(2)_H$ symmetry and Lorentz invariance. 
In other words, an inert Higgs doublet $H_2$ emerges naturally in G2HDM 
without resorting to the discrete $Z_2$ symmetry~\footnote{After symmetry breaking in G2HDM, one can actually show that an effective $Z_2$ symmetry 
emerges~\cite{private}.}.
We specify the most general and renormalizable scalar potential invariant under
$SU(2)_L \times U(1)_Y \times SU(2)_H \times U(1)_X$ in Appendix~\ref{section:app} and in turn
discuss the scalar mass spectra in Appendix~\ref{section:app_spectrum}.

To generate masses for the SM fermions via Yukawa couplings in an $SU(2)_H$ invariant manner, we choose to pair SM right-handed fermions with new right-handed
ones into $SU(2)_H$ doublets, whereas the SM left-handed fermions are singlets under $SU(2)_H$ 
as indicated in Table~\ref{tab:quantumnos}.
In addition, to make all new fermions massive via the vev of the $SU(2)_H$ doublet $\Phi_H=(\Phi_1,\Phi_2)^T$, extra left-handed fields
$f^H_L$~($f=d,u,e,\nu$) are introduced. The corresponding $SU(2)_H$ invariant 
Yukawa couplings are
\begin{align}
\mathcal{L}_{\rm Yuk} \supset \, & -  y'_d \overline{d^H_L} \lee d^H_R \Phi_2 - d_R \Phi_1  \rii 
- y'_u \overline{u^H_L} \lee u_R \Phi^*_1 + u^H_R \Phi^*_2  \rii   \nn \\
& -  y'_e \overline{e^H_L} \lee e^H_R \Phi_2 - e_R \Phi_1  \rii   
- y'_\nu \overline{\nu^H_L} \lee \nu_R \Phi^*_1 + \nu^H_R \Phi^*_2 \rii+ {\rm H.c.}\, .
\label{eq:YuK_new} 
\end{align}
With a non-vanishing $\lan \Phi_2 \ran$, the four Dirac fields $d^H$,
$u^H$, $e^H$ and $\nu^H$ acquire a mass of 
$y'_d \lan \Phi_2 \ran$, $y'_u \lan \Phi_2 \ran$, $y'_e \lan \Phi_2 \ran$, and $y'_\nu \lan \Phi_2 \ran$,
respectively.
On the other hand, the SM quarks and leptons obtain their masses
from the vev of $H_1$ via the Yukawa couplings 
\begin{align}
\mathcal{L}_{\rm Yuk} \supset & 
 + y_d \bar{Q}_L \lee d^H_R H_2 - d_R H_1  \rii - y_u \bar{Q}_L \lee u_R \tilde{H}_1 + u^H_R \tilde{H}_2  \rii  \nn\\
& +\, y_e \bar{L}_L \lee e^H_R H_2 - e_R H_1 \rii 
- y_\nu \bar{L}_L \lee \nu_R \tilde{H_1} + \nu^H_R \tilde{H_2} \rii  + {\rm H.c.} \, ,
\label{eq:Yuk_SM}
\end{align}
with $\tilde H_{1,2} = i \tau_2 H_{1,2}^*$. 
Note that in both Yukawa couplings given in (\ref{eq:YuK_new}) 
and (\ref{eq:Yuk_SM}) only the SM Higgs doublet $H_1$ couples bilinearly with the SM fermions. 
Thus FCNC interactions for the SM fermions are absence at tree level naturally in G2HDM.
It also implies the new heavy fermions can decay into SM fermions plus DM via the Yukawa couplings in (\ref{eq:Yuk_SM}). 
For instance, $f^H_R \to f_L H_2^{0*}$ where $H_2^{0*}$ is a DM candidate and 
manifests as the missing transverse energy. 
We note that absence of FCNC interactions in 2HDM by embedding the discrete $Z_2$ symmetry 
into an extra $U(1)^\prime$ has been studied in~\cite{Ko:2012hd,Ko:2013zsa,DelleRose:2017xil,Campos:2017dgc}.

There are $SU(2)_H$ gauge bosons, $W^{\prime (p,m) }$ and $W^{\prime 3 }$, and the $U(1)_X$ gauge boson $X$, apart from the SM ones.
Due to the symmetry breaking pattern, $W^{\prime (p,m) }$ will not mix with the SM counterparts but $W^{\prime 3 }$ and $X$ mix with
the SM $SU(2)_L$ $W^3$ and $U(1)_Y$ $Y$ gauge boson.
In this setup, besides the SM massless photon corresponding to the unbroken generator 
$Q= T^3_{L} + Y$, there exists a massless dark photon 
corresponding to the unbroken generator $Q_D = 4 \cos^2 \theta_w T^3_{L} - 4 \sin^2 \theta_w Y 
+ 2 T^3_{H} + X$. Here $T^3_{L}$ ($T^3_{H}$) is the third generator of $SU(2)_L$~($SU(2)_H$), 
$Y(X)$ is the $U(1)_Y$~($U(1)_X$) generator, and $\theta_w$ is the Weinberg angle.
Such a massless dark photon could be cosmologically problematic.
To circumvent the problem, one can 
resort to the Stueckelberg mechanism to give a mass to the $U(1)_X$ gauge boson as in~\cite{Stueckelberg}. 
One could take this mass to be large enough so that
$X$ is decoupled from the particle spectra.
Another way out is to treat $U(1)_X$ as a global symmetry as was proposed in~\cite{DiazCruz:2010dc} and 
adopted in~\cite{Huang:2015wts} as well. We will follow the same strategy in what follows.

In this case, after diagonalizing the mass matrix of $Y$, $W^3$
and $W^{\prime 3}$, one obtains massless $\gamma$, massive $Z$ and $Z^\prime$.
Furthermore, the mixing between $Z - Z^\prime$ is constrained to be of order $10^{-3}$ for TeV $Z^\prime$ because of the electroweak precision measurements~\cite{Erler:2009jh}. As a consequence, impacts of the mixing are numerically negligible and will be ignored. The resulting $SU(2)_H$ gauge boson mass spectrum is
\begin{align}
m^2_{W^{\prime\, (p,m)}}  & = \frac{1}{4} g^2_H \lee v^2 + v^2_\Phi + 4 v^2_\De \rii \, , \nn\\
m^2_{Z^\prime} & =  \frac{1}{4} g^2_H  \lee v^2 + v^2_\Phi \rii \, ,
\label{eq:zw_mas}
\end{align}
where $ (v/\sqrt{2}, \, v_\Phi/\sqrt{2} , \, - v_\De  ) = ( \lan H^0_1 \ran,  \, \lan \Phi_2 \ran, \, \lan \De_3 \ran ) $.
Note that $W^{\prime\, (p,m)}$ is always heavier than $Z^\prime$ in G2HDM.

As the SM right-handed fermions as well as the new fermions are charged under $SU(2)_H$, they couple to the $W^{\prime\, (p,m)}$ and $Z^\prime$ bosons.
The relevant gauge interactions without the $Z-Z^\prime$ mixing read  
\begin{align}
{\cal L} \supset  {\cal L}(W) +  {\cal L}(\ga) + \Delta {\cal L} \, .
\end{align}
Here ${\cal L}(W)$ and ${\cal L}(\ga)$ refer to the charged current mediated by the $W$ boson and the electric current by the photon $\ga$ respectively, 
\begin{align}
{\cal L}(\ga) &= \sum_f Q_f e \bar{f} \ga^\mu f A_\mu \,  , \nn \\
{\cal L}(W) &= \frac{g}{\sqrt{2}} \lee \overline{\nu_L} \ga^\mu e_L + \overline{u_L} \ga^\mu d_L \rii W^+_\mu + {\rm H.c.} \, ,
\end{align}
where $Q_f$ is the corresponding fermion electric charge in units of $e$. 
$\Delta {\cal L}$ represents (electrically) neutral current interactions of the massive 
bosons, $Z$, $Z^\prime$ and $W^{\prime (p,m)}$~(for demonstration,
only the lepton sector is shown but it is straightforward to include the quark sector): 
\begin{equation}
\Delta {\cal L} =  {\cal L}(Z) +{\cal L}(Z^\prime) +  {\cal L}(W^{\prime (p,m)}) \, ,
\end{equation}
where 
\begin{align}
 {\cal L}(Z) & =  \frac{g}{\cos\th_w}  J^\mu_{\tiny{Z}}   Z_\mu \,\, , \nn\\
  {\cal L}(Z^\prime) &=  g_H   J^\mu_{W^{\prime 3}}   Z^{\prime}_\mu  \,,   \\
 {\cal L}(W^{\prime (p,m)}) &= \frac{1} {\sqrt{2}} g_H \left( 
 J^\mu_{W^{\prime m}} W^{\prime p}_\mu + {\rm H.c.}  \right) \,, \nn
\end{align}
and
\begin{align}
J^\mu_{\tiny{Z}} & = 
 \sum_{f= e,\nu} \lee \overline{f_L} \ga^\mu (T^3_L - Q_f \sin^2\th_w)  f_L + \overline{f_R} \ga^\mu (- Q_f \sin^2\th_w)  f_R \rii
+ \sum_e \overline{e^H_R} \ga^\mu ( \sin^2\th_w) e^H_R  \,, \nn\\
J^\mu_{W^{\prime 3}} & = \sum_{f_R=N_R, E_R}  \overline{f_R} \ga^\mu (T^3_H)  f_R \,, \\
J^\mu_{W^{\prime m}} & = \sum_e \lee \overline{e^H_R} \ga^\mu e_R + \overline{\nu_{eR}} \ga^\mu \nu^H_{eR}  \rii  \, . \nn
\end{align}
The current interactions in ${\cal L}(W^{\prime (p,m)})$ and 
${\cal L}(Z^\prime)$ will dictate how
$W^{\prime\, (p,m)}$ and $Z^\prime$ decay into SM and heavy fermions, 
and determine which final states one should look into for collider searches.

\section{Methodology}
\label{sec:methodology}

To simulate the total cross sections and various distributions for the relevant processes in the colliders, 
we will follow the standard protocol well 
established by many collider phenomenologists.
We use \texttt{FeynRules}~\cite{Alloul:2013bka} to build up the model files 
for G2HDM and pass it to \texttt{Madgraph5}~\cite{Alwall:2014hca} 
for the matrix element calculation 
and event generation. We simulate parton showering by using \texttt{Pythia8.1}~\cite{Sjostrand:2007gs}, 
and employ \texttt{Delphes3}~\cite{deFavereau:2013fsa} for detector simulations. 
Finally, the package \texttt{MadAnalysis5}~\cite{MA5} is used to analyze the simulation data.

In the G2HDM, apart from the extra gauge bosons $W^{\prime (p,m)}$ and $Z'$, additional heavy fermions have to be included to
attain gauge invariant Yukawa couplings as explained above. 
To simplify the analysis, we assume two universal masses for the heavy fermions, one for leptons and the other for quarks. 
As a result, there are five relevant mass scales in our analysis, namely the masses of the dark matter particle $H^{0*}_2$, 
the heavy leptons $L^{H}=(e^H, \mu^H, \tau^H)$ and $\nu^{H}=(\nu^{H}_e, \nu^H_{\mu}, \nu^H_{\tau})$, the heavy quarks $Q^{H}=(u^H,d^H,c^H,s^H,t^H,b^H)$, and the two heavy gauge bosons~$W^{\prime (p,m)}$ and $Z'$.
In addition, the new charged fermions have to be heavier than 100 GeV, 
a constraint inferred from the combined analysis of the LEP2 run data by the four LEP collaborations~\cite{LEP2}.
We will study the following two benchmark mass spectra for the new fermions,
while the new gauge bosons are always assumed to be heavier than $1.5\,\tev$ 
such that the gauge coupling $g_H$ is not too small.
\begin{description}

\item[Spectrum-A] Heavy and decoupled new quark scenario.\\
The new quarks $Q^{H}$ are chosen to be heavier 
 than $Z'$. Specifically we take $ m_{Q^{H}}=m_{Z'}+1$ TeV, and thus
 channels of the new quarks will not be considered in the $Z'$-resonance searches. 
 On the other hand, the new leptons $L^{H}$($\nu^{H}$) are assumed to be lighter than $Z^\prime$ 
with $ m_{L^{H}(\nu^{H})}=2\, m_D$ in which $m_D$ is the dark matter mass. Hence $L^H$($\nu^{H}$) can be pair 
produced by $Z'$ on-shell decays. 
In order to well separate the spectrum, 
we fix the mass ratio between DM and the $SU(2)_H$ gauge bosons:  
 $ m_{Z'}\simeq m_{W^{\prime (p,m)}}=5 \,m_D$.
 
 \item[Spectrum-B] Light new quark scenario. \\
 For completeness, we also study a scenario with lighter new quarks where the new heavy quarks and leptons are degenerate: 
 $ m_{Q^{H}}=m_{L^{H}(\nu^{H})}=2 \, m_D $, while the same DM-$Z^\prime(W^{\prime (p,m)})$ mass ratio 
 $ m_{Z'}\simeq m_{W^{\prime (p,m)}}=5 \, m_D$ as in Spectrum-A is assumed.
 
 \end{description}
 
To achieve $m_{Z'}\simeq m_{W'}$, one needs $v_\Phi \gtrsim 3 v_\Delta \gg v$ based on Eq.~\eqref{eq:zw_mas}. Note that this setup is different from the previous work~\cite{Huang:2015wts},
 where $v_\Delta \gtrsim v_\Phi \gg v$ was assumed. 
Furthermore, for simplicity decays of the heavy gauge bosons into scalar Higgs pairs
are presumed to be either kinematically forbidden or negligible. It is justified since all of the new scalars except for 
DM can be heavier than $W'$ and $Z'$ as displayed in the last table in
Appendix~\ref{section:app_spectrum}.
Moreover, the coupling between the longitudinal components of $W'$ and $Z'$ and 
the DM can in principle be made small by varying the parameters in the scalar potential.
In this way, the transverse components~(whose coupling to DM is simply $g_H$) 
govern decays of  $W'$ and $Z'$ into DM particles
but this  contribution to the decays is subleading compared to those of 
the heavy fermions in the final states, given the larger number of the new fermions in the model. 

In both scenarios, 
the new heavy fermions are kinematically allowed to be produced by 
either $Z^\prime$ or $W^{\prime (p,m)}$ decays.
As a result, we propose searches for the new fermions as follows.~\footnote{
The symbol $l$ refers to the first and second generation charged leptons, $e$ and $\mu$, as well as their antiparticles,
while $\tau$ denotes the third generation ones.
Similarly, $j$ refers to a light quark/anti-quark jet of the first and second generations, 
and $b$ and $t$ are the bottom/anti-bottom and top/anti-top jets respectively.}
 
\begin{itemize}

\item For Spectrum-A, the heavy charged leptons can be produced via 
$p p\rightarrow Z' \rightarrow L^H \overline{L^H}$ and 
$p p\rightarrow W'^p W'^m \to \overline{L^H} L \overline{L} L^H$,
and the corresponding final states will be (1) 2$ l +\mET$, (2) 2$ \tau +\mET$, (3) 4$ l +\mET$, 
(4) 2$ l $+2$ \tau+\mET $, and (5) 4$ \tau +\mET$.

\item  For Spectrum-B, the new quark pairs can also be on-shell produced through
$p p\rightarrow Z' \rightarrow Q^H \overline{Q^H}$, and 
thus the following final states (1) 2$ j +\mET$, (2) 2$ b +\mET$, 
and (3) 2$ t +\mET$ will be considered.
These processes are relevant to the dijet plus missing transverse energy searches for $Z^\prime$. 
Needless to say, the continuum contributions from QCD to the new quark pair 
production should be taken into account.

\end{itemize}

\begin{table}[hb!]
\caption{\small  \label{tab:BR-Zp}
Branching ratios for different decay modes of $Z'$ with $  1.5 \leq m_{Z'} \leq 3$ TeV.
Here $Q$ denotes 6 quark flavors ($u, d, c, s, t, b$) and $L$~($\nu$) represents 3 lepton flavors ($e~(\nu_e), \mu~(\nu_\mu) , \tau~(\nu_\tau)$).
}
\vspace{1.0mm}
\begin{ruledtabular}
\begin{tabular}{ l c c c c c c }
$Z'$ & $ BR(Q\overline{Q}) $ & $ BR(L^+L^-) $ & $ BR(\nu\overline{\nu}) $ & $ BR(Q^H\overline{Q^H}) $ & $ BR(L^H\overline{L^H}) $ & $ BR(\nu^H\overline{\nu^H}) $ \\ \hline
Spectrum-A & $ 66.52\% $ & $ 11.13\% $ & $ 11.13\% $ & -- & $ 5.61\% $ & $ 5.61\% $ \\ 
Spectrum-B & $ 49.84\% $ & $ 8.31\% $ & $ 8.31\% $ & $ 25.14\% $ & $ 4.20\% $ & $ 4.20\% $\\
\end{tabular}
\end{ruledtabular}
\end{table}

\begin{table}[hb!]
\caption{\small  \label{tab:BR-Wp}
Branching ratios for different decay modes of $W^{\prime (p,m)}$ with $  1.5 \leq m_{W^{\prime (p, m)}} \leq 3$ TeV.
}
\vspace{1.0mm}
\begin{ruledtabular}
\begin{tabular}{ l c c c }
$W^{\prime (p,m)}$ & $ BR(Q^H\overline{Q},Q\overline{Q^H}) $ & $ BR(L^H\overline{L}, L\overline{L^H} ) $ & $ BR(\nu^H\overline{\nu},\nu\overline{\nu^H} ) $ \\ \hline
Spectrum-A & -- & $ 50\% $ & $ 50\% $ \\ 
Spectrum-B & $ 74.96\% $ & $ 12.52\% $ & $ 12.52\% $\\
\end{tabular}
\end{ruledtabular}
\end{table}

In Table~\ref{tab:BR-Zp} and \ref{tab:BR-Wp}, we list the branching ratios for the $Z'$ and $W^{\prime (p,m)}$ 
decays respectively in the two scenarios.\footnote{
In fact, the branching ratio for each decay mode is insensitive to $ m_{Z'} $ and 
$ m_{W^{\prime (p,m)}}$ in the region of interest from $ 1.5$ to $3$ TeV.} 
The $Q^H\overline{Q^H}$ final state in the $Z^\prime$ decay
is kinematically allowed in Spectrum-B, resulting in smaller partial decay widths
into the SM $Q\overline{Q}$ and $L^+L^-$ final states compared to Spectrum-A. 
On the other hand, since $W^{\prime (p,m)}$ do not decay into SM fermion pairs,
the opening of $Q^H\overline{Q}+Q\overline{Q^H}$ final state in Spectrum-B affects only the other exotic leptonic channels.
These exotic decays can be phenomenologically interesting 
as we will see in the next section.

Note that $Z^\prime$ and $W^{\prime (p,m)}$ 
can also decay into scalars, $H$, $\Phi_H$
and $\De_H$, which are charged under $SU(2)_H$.
The branching fractions, however,
depend on the scalar mixing parameters in the scalar potential~\cite{Huang:2015wts} 
which can make the analysis rather convoluted.
As mentioned earlier, to simplify the analysis in this work, we neglect scalar final states 
and instead focus on the fermion channels at which the corresponding partial decay widths
are simply fixed by the $SU(2)_H$ gauge coupling as well as the new heavy fermion masses.

\section{$Z'$ Searches At The LHC}
\label{sec:Zprime}

In this section, we first present the $Z'$ constraints, derived
from the ATLAS and CMS dijet and dilepton searches based on the $13\,\tev$ data. Then we propose potential 
$Z'$ signatures from exotic decay searches which have smaller cross sections than direct $Z^\prime$ searches but can be explored at  the $14\,\tev$ HL-LHC.
In Section~\ref{sec:Wprime}, we will investigate the $W^{\prime (p,m)}$ searches at a 
future proton-proton 100 TeV collider. 
As mentioned before, unlike $Z^\prime$ which can be singly created and  probed directly by dilepton and dijet searches at the LHC, 
the heavier $W^{\prime (p,m)}$ must be produced in pair in light of the $SU(2)_H$ gauge symmetry and therefore are less constrained.

\subsection{Constraints on $Z'$ from current dilepton and dijet searches}

In G2HDM, $Z'$ is always lighter than 
$W'$ and can be directly probed by resonance searches as mentioned above.
Due to null results of the direct searches, very stringent limits 
are imposed on any model of $Z^\prime$ that directly couples to SM fermions. 
For instance, the sequential SM with $Z^\prime$ having the same couplings to SM fermions 
as the SM $Z$ gauge boson is constrained to be heavier than $4$ TeV~\cite{ATLAS:2016cyf, CMS:2016abv}.

\begin{figure}[th!]
\centering
\includegraphics[width=3.5in, height=3.5in]{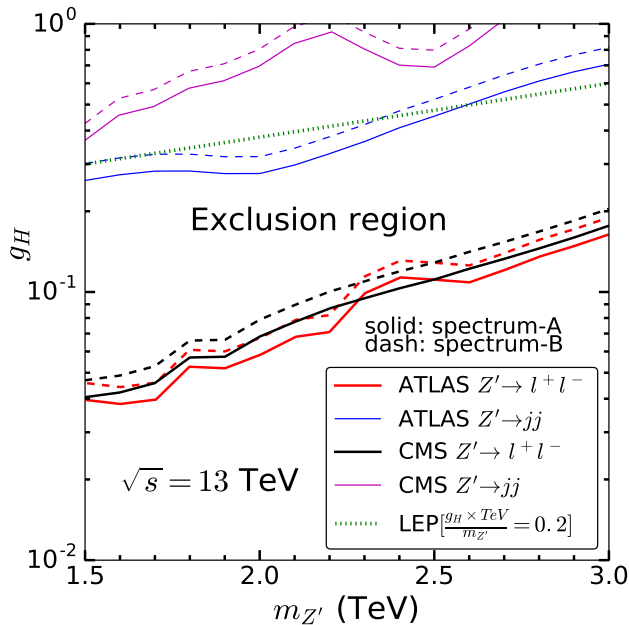}
\caption{\small \label{fig:zp_con}
The $Z'$ constraints for G2HDM inferred from the latest ATLAS and CMS $13\tev$ results.  
The solid lines denote Spectrum-A while the dashed lines 
refer to Spectrum-B. The main differences between the two scenarios are the branching fractions 
of $Z^\prime$ into the SM quarks and leptons, as shown in Table~\ref{tab:BR-Zp}.
}
\end{figure}

Recently, ATLAS and CMS collaborations have reported their 
updated results of $Z'$ resonance searches for channels of
dilepton~\cite{ATLAS:2016cyf, CMS:2016abv}, dijet~\cite{ATLAS:2016lvi, CMS:2016wpz}, $b$-quark pair~\cite{ATLAS:2016gvq}, $t$-quark pair~\cite{TheATLAScollaboration:2016wfb}, and other bosonic final states~\cite{ATLAS:2016kxc, ATLAS:2016cwq, ATLAS:2016yqq} at 13 TeV. 
In light of the irreducible QCD background at the LHC, 
the dilepton channel is the cleanest one to reconstruct the invariant mass of the final state 
particles and yields the most stringent constraints. 
In this work, we consider two major type of constraints: dilepton and dijet channels. 
We calculate the cross sections of $p p\rightarrow Z' \rightarrow l^{+}l^{-}/ j j $ 
with the help of \texttt{Madgraph5}~\cite{Alwall:2014hca} and 
compare them with the latest constraints from the LHC. 

In Fig.~\ref{fig:zp_con}, we present the exclusion regions of $g_H$ as a function of $m_{Z^\prime}$. 
The solid lines correspond to Spectrum-A and the dashed lines denote Spectrum-B.
The major discrepancies between the two scenarios are the branching ratios of the $Z^\prime$ decay 
into SM quarks and leptons, as shown in Table~\ref{tab:BR-Zp}.
Compared with the previous constraints~\cite{Huang:2015wts} obtained from the 8 TeV data, 
the improvement is about a factor of two in the region of $ 1.5 <m_{Z^\prime} < 2.25$ TeV. 
In addition, thanks to the higher center-of-mass energy $\sqrt{s}$ ($8 \to 13$ TeV), 
the bounds for $m_{Z'}>2.25\, \tev$ are significantly improved
and become stronger than the LEP limits based on the $\sigma(e^+ e^- \to l^+ l^-)$ measurements~\cite{Huang:2015wts}. 

We note that the constraints on the $Z^\prime$ mass in G2HDM 
is less stringent than the $Z'$ in LRSM or LRTHM where a discrete symmetry is imposed
to equate the new gauge coupling to the SM $SU(2)_L$ one. 
The price to pay for G2HDM is of course a smaller $g_H$.

\subsection{$Z'$ exotic decays into heavy fermions}

We now move on to the $Z'$ exotic decays which can shed light on the existence of exotic fermions in G2HDM.      
As noted before, the scalar decay channels of $Z^\prime$ depending on details of the complicated
scalar potential and hence are ignored in this work.
On the other hand, the heavy fermion channels, which are governed by $g_H$ and the heavy fermion
masses only, can be easily addressed.

Thinking forwardly and optimistically, one can envisage that a $Z^\prime$ will be 
discovered by the direct searches of 
dilepton and dijet channels in the foreseeable future at HL-LHC. 
If so, the heavy fermions in G2HDM can also be probed via $Z^\prime$ 
on-shell decays if kinematically allowed.
In order to perform a more general study for this purpose, we will temporarily 
relax the mass relations among the heavy fermions, $Z^\prime$, and DM
for both Spectrum-A and Spectrum-B mentioned in Section~\ref{sec:methodology}.
To be specific, in this section and only in this section, 
we will relax the fixed mass relation in Spectrum-A to
$ 2\,m_{L^H} < m_{Z^\prime} < 2\,m_{Q^H}$, 
and for Spectrum-B we will assume $ 2\,m_{ ( L^H,Q^H ) } < m_{Z^\prime}$ instead.
Besides, as long as the mass differences between the heavy fermions and DM are large enough, 
the actual value of $ m_{\rm DM} $ will not have a significant impact on the analysis. Thus we will 
choose a nominal value of $m_{\rm DM} = 50$ GeV in the following analysis. 
These two modified spectra will be referred as Spectrum-A$^\prime$ and Spectrum-B$^\prime$ in what follows.

Owing to the $ SU(2)_H $ symmetry, final states of the exotic decay modes 
are quite similar to those used to search for supersymmetric (SUSY) particles 
with R-parity conservation in the context 
of Minimal Supersymmetric Standard Model (MSSM).
For example, for the slepton ($\tilde{l}^{\pm}$) searches at the LHC the major process is
$ 2l +\mET $ channel, namely
\begin{equation}
p p\rightarrow \gamma /Z \rightarrow \tilde{l}^{+}\tilde{l}^{-} \rightarrow 2l +\mET \; .
\label{eq:slepton}
\end{equation}
Similarly, in G2HDM decays of $Z^\prime$  into a pair of exotic fermions can also lead to 
the same final states:
\begin{equation}
p p\rightarrow Z' \rightarrow l^H\overline{l^H} \rightarrow 2l +\mET \;.
\label{eq:lH}
\end{equation} 
These two processes with the same final states exhibit analogous event topology, 
allowing us to apply the same kinematic cuts.~\footnote{
The process in Eq.~\eqref{eq:lH} is the major contribution to the $ 2l +\mET $ channel in G2HDM, although there are also the processes, $ p p\rightarrow \gamma /Z \rightarrow l^H\overline{l^H} \rightarrow 2l +\mET $, similar to those in Eq.~\eqref{eq:slepton}.
}
We calculate the cross sections for the $Z'$ exotic decays and then impose bounds from SUSY searches on these decays.

The bounds on the heavy fermion masses can be mitigated in a scenario of 
the compressed mass spectrum: $m_{  (L^H , Q^H ) } \gtrsim m_{\rm DM}$. 
In this case mono-X~($X= \ga$, $g$, $W$, $Z$$\cdots$) + $\mET$, in particular the mono-jet + $\mET$ signal, 
can be used to search for DM  as in the MSSM with the compressed mass spectrum.
This scenario, however, will not be considered here 
as the mass difference $m_ {(L^H , Q^H)} - m_{\rm DM}$ is taken to be large.

\begin{figure}[th!]
\centering
\includegraphics[width=3.in, height=3.in]{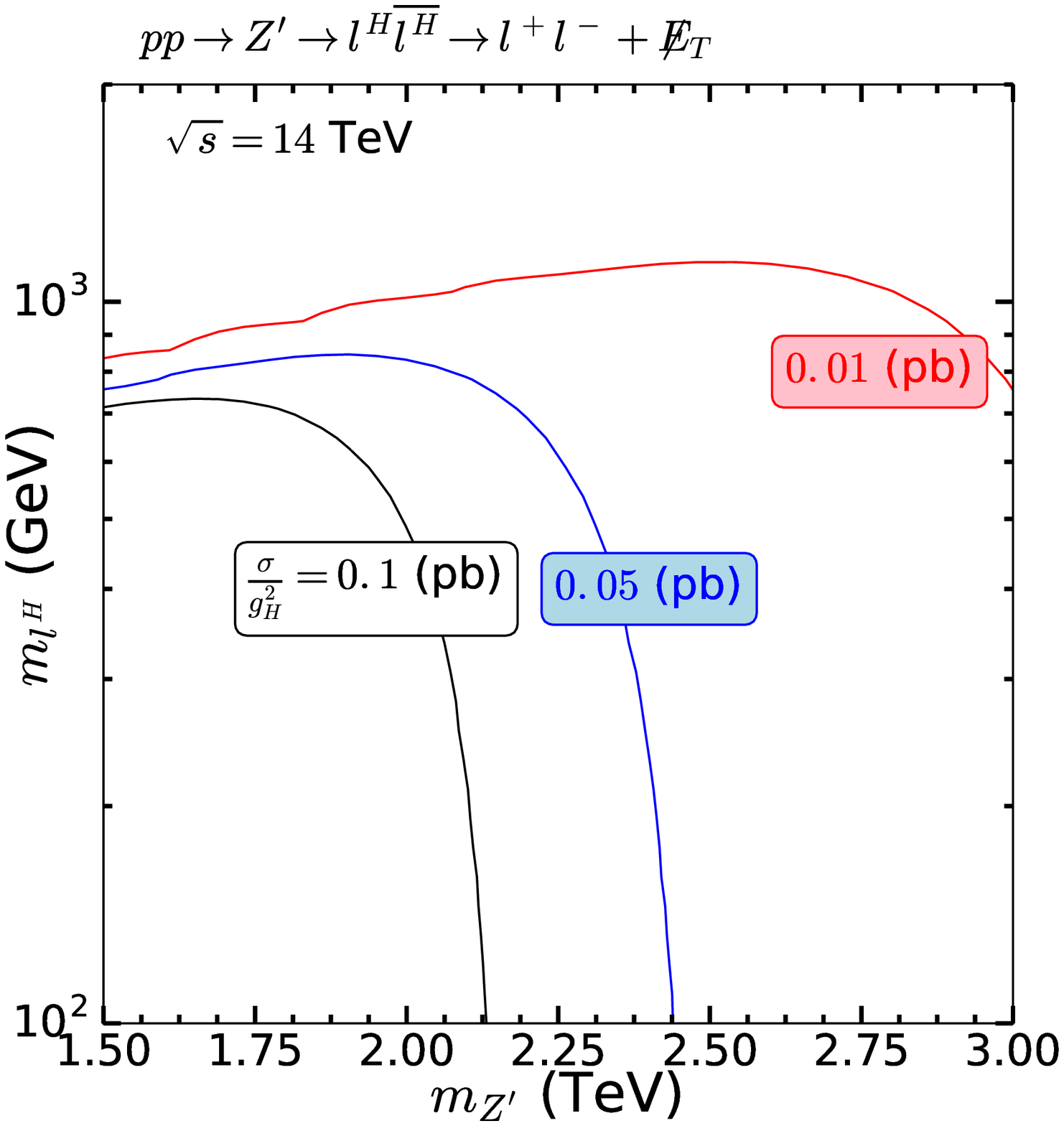}
\includegraphics[width=3.in, height=3.in]{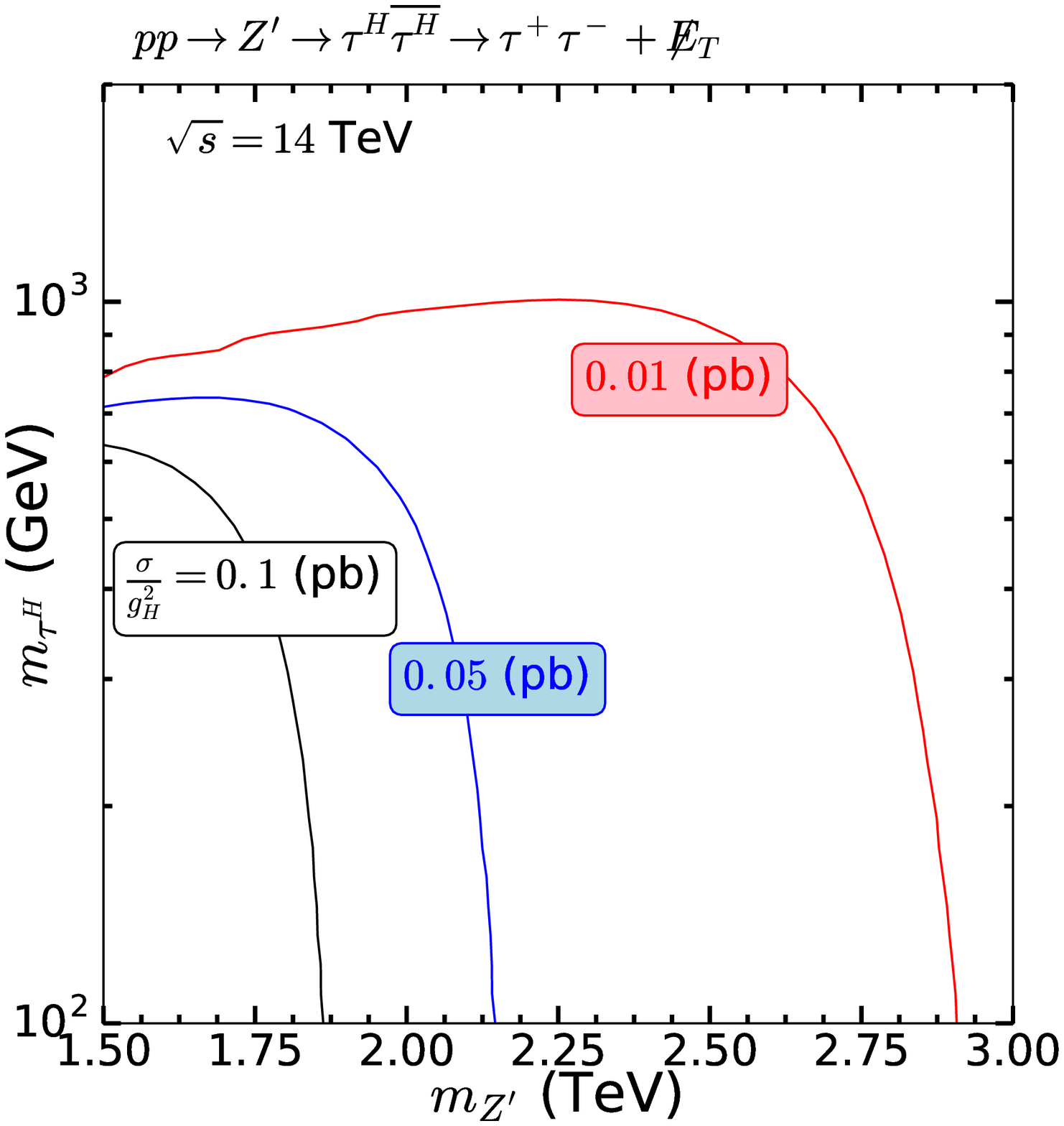}
\caption{\small \label{fig:SPA}
The cross section contours for Spectrum-A$^\prime$ at $\sqrt{s}=14\, \tev$  on the planes of  
($ m_{Z'} $, $ m_{l^H} $) and ($ m_{Z'} $, $ m_{\tau^H}$). 
Three benchmark values are shown here: $\frac{\sigma}{g^2_H}=0.1$~(black), $0.05$~(blue), 
and $0.01$~(red) in units of pico-barn (pb).
}
\end{figure}


We now present the constraints on the $Z^\prime$ exotic decays. 
With the modified spectra, we will be able to obtain contours of the 
production cross sections on the $(m_{Z^\prime}, m_{f^H})$ plane and 
compare with the LHC limits on the SUSY particle searches.

\begin{itemize}

\item

Spectrum-A$^\prime$: $ 2\,m_{L^H} < m_{Z^\prime} < 2\,m_{Q^H}$ 
and $m_{\rm DM} = 50$ GeV. 

We concentrate on the following two channels,  
\begin{align}
\label{eq:zp_exo_A}
p p\rightarrow Z' \rightarrow l^H\overline{l^H} \rightarrow 2l +\mET \,,  \\
p p\rightarrow Z' \rightarrow \tau^H\overline{\tau^H} \rightarrow 2\tau +\mET \,,
\label{eq:zp_exo_B}
\end{align}
where $ l^H = (e^H, \mu^H) $, and $ l = (e, \mu) $. 
In Fig.~\ref{fig:SPA},
the contour plots for the cross sections of the processes in~\eqref{eq:zp_exo_A} 
and \eqref{eq:zp_exo_B}
are shown on the planes of $ m_{Z'} - m_{l^H} $ (left panel) and $ m_{Z'} - m_{\tau^H} $ (right panel).
Since the cross section is proportional to $g^2_H$ for on-shell heavy fermions\footnote{
For most of the regions of interest, the extra heavy fermions 
from $Z^\prime$ decays are on-shell.},
the results are shown in terms of $\sigma/g^2_H$ to factor out the $g_H$ dependence.
For a specific value of $g_H$, one can simply rescale the contours by $g^2_H$
whose limits for a given value of $m_{Z^\prime}$ have been presented in Fig.~\ref{fig:zp_con}.
The black, blue, and red contours correspond to $\sigma / g^2_H=0.1$, $0.05$, 
and $0.01$ pico-barn~(pb) respectively, assuming $\sqrt{s}=14$ TeV. 

We employ the recent results of ATLAS SUSY searches for neutralinos and charginos
based on the 2$ l +\mET$ and 2$ \tau +\mET$ channels 
to constrain the $Z^\prime$ exotic decays in G2HDM.
The resulting bounds should be regarded as
estimated constraints, as the signal regions and efficiency may have some differences between MSSM and G2HDM.
For the left panel of Fig.~\ref{fig:SPA} ($2l+\mET$ channel), 
we use the signal region \texttt{SR2l-A}  which refers to a set of event selections 
listed in Table 1 of Ref.~\cite{ATLAS:2016uwq}.  
It gives rise to the constraint $ \langle\epsilon\sigma\rangle^{95}_{{\rm obs}}\leq 1.89 $ fb at 13 TeV. 
Assuming that the factor of signal efficiency $ \epsilon $ is of $ \mathcal{O}(1) $ at 14 TeV
\footnote{ 
If the magnitudes of the cross sections at 14 TeV are just slightly larger than those at 13 TeV, the constraints we present here will be more stringent than the 13 TeV ones since the detection efficiency of order 1 has been assumed.
}, 
we can infer limits on $g_H$ at the 14 TeV LHC in the following way.
For instance, to satisfy the \texttt{SR2l-A} bound $\sigma < 1.89$ fb, 
along the black, blue and red contours in the left panel of Fig.~\ref{fig:SPA}, 
the corresponding $g_H$ is required to be smaller than $0.137$, $0.194$ and $0.435$, respectively.

Likewise, for the $2\tau +\mET$ channel on the right panel of Fig.~\ref{fig:SPA},
we utilize the \texttt{SRC1C1} signal region in Ref.~\cite{ATLAS:2016ety}. 
It yields $ \langle\epsilon\sigma\rangle^{95}_{{\rm obs}}\leq 0.33 $ fb at 13 TeV  
that demands $ g_H $ to be less than $0.057 $, $0.081 $, and $0.182$
for the black, blue, and red contours, respectively.

\begin{figure}[th!]
\centering
\includegraphics[width=2.65in, height=2.65in]{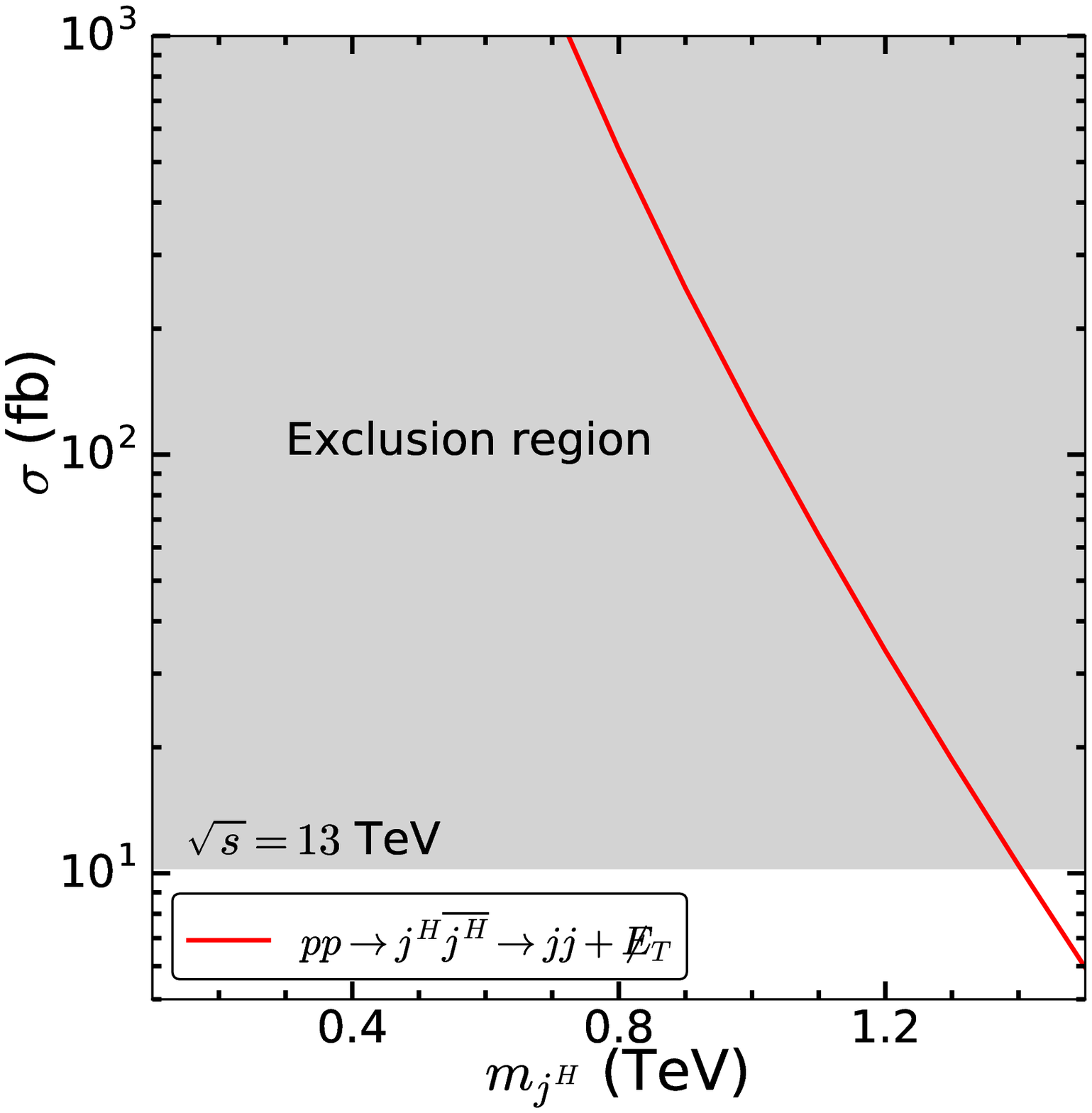}
\includegraphics[width=2.65in, height=2.65in]{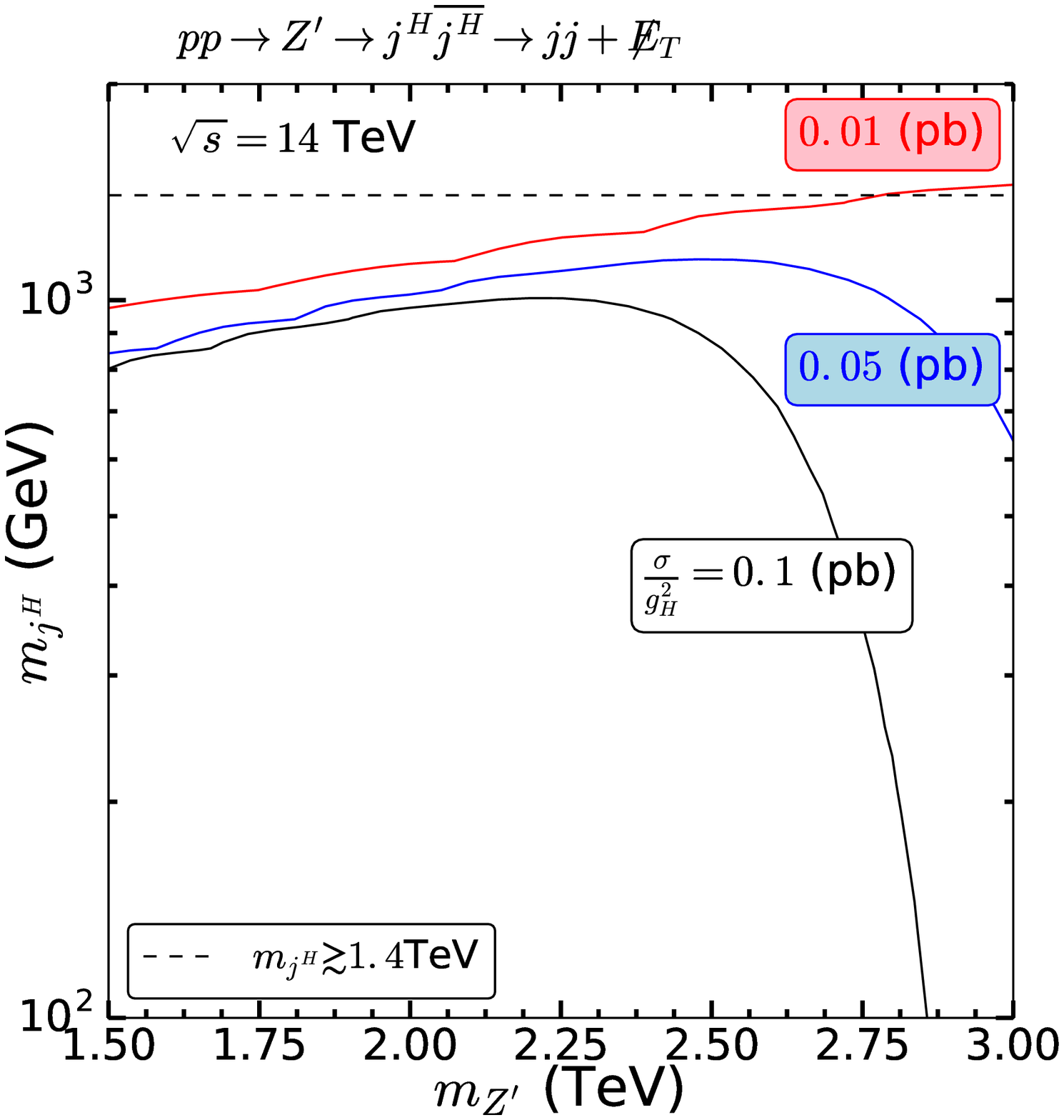}\\
\includegraphics[width=2.65in, height=2.65in]{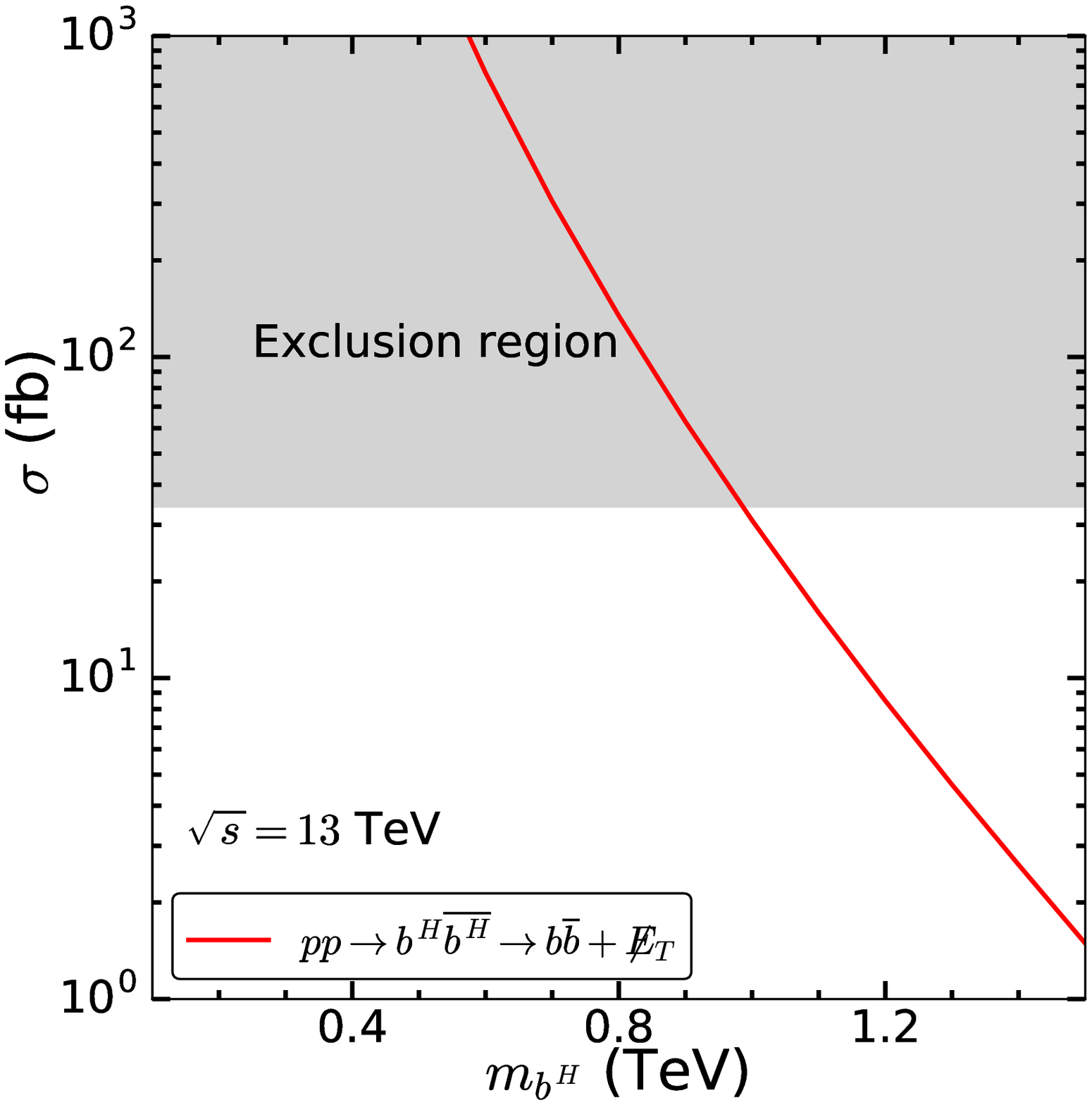}
\includegraphics[width=2.65in, height=2.65in]{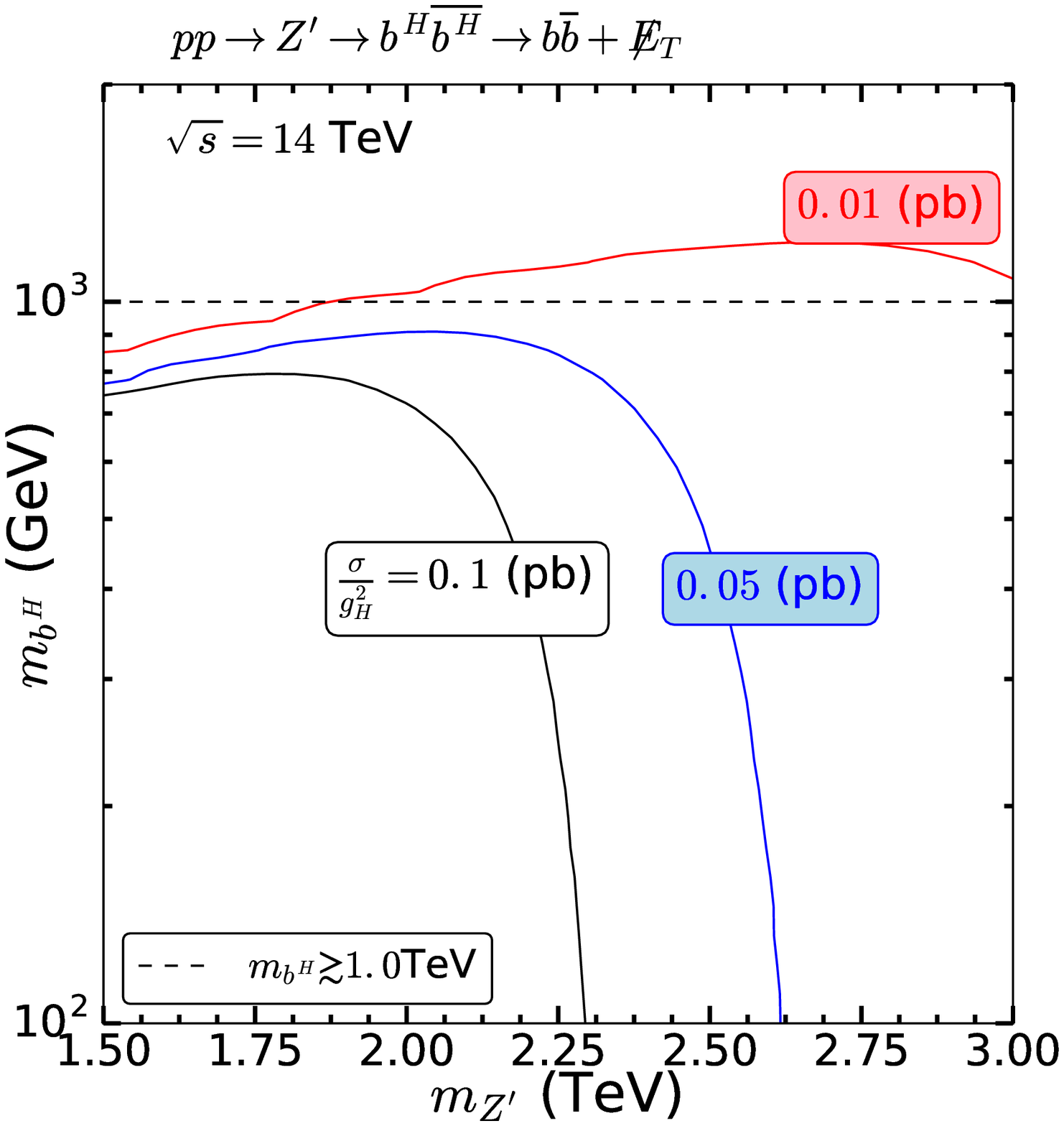}\\
\includegraphics[width=2.65in, height=2.65in]{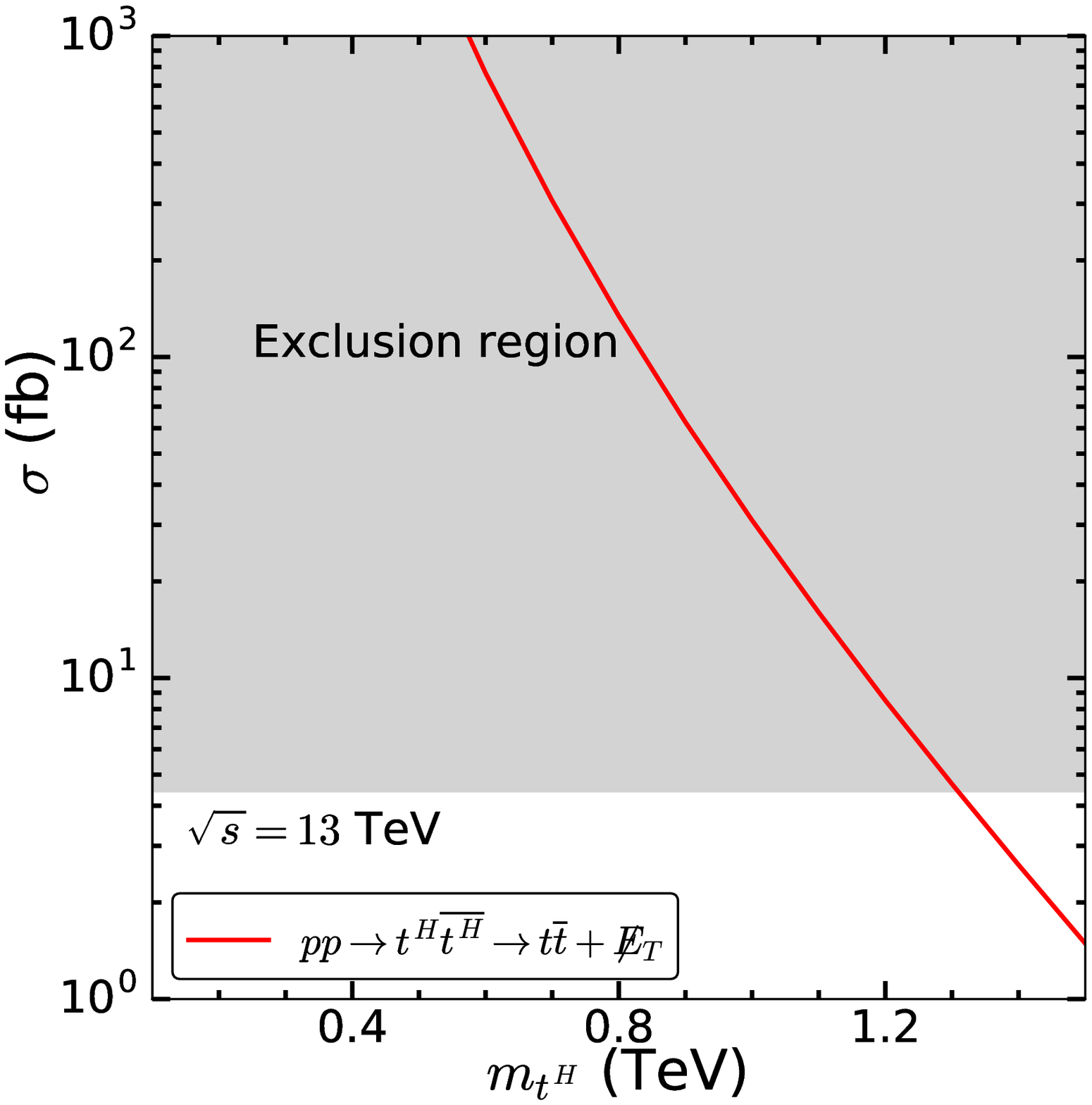}
\includegraphics[width=2.65in, height=2.65in]{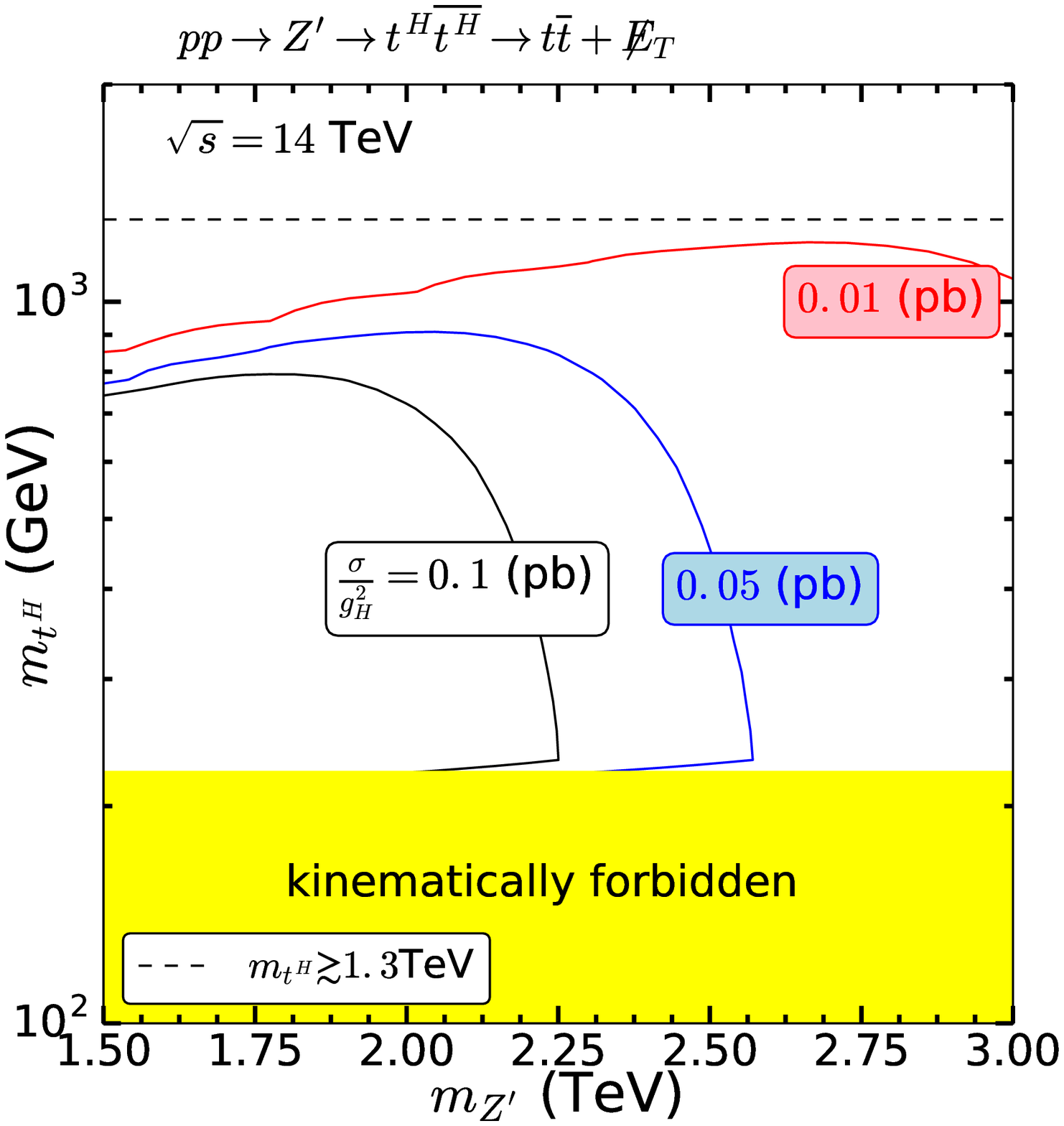}
\caption{\small \label{fig:SPB}
The new quark searches in Spectrum-B$^\prime$. 
Left column: the cross sections for $ 2j+\mET $, $ b\overline{b}+\mET $, and $ t\overline{t}+\mET $ channels 
via the important strong interactions at $\sqrt{s}=13\, \tev$  as functions of $m_{Q^H}$. The gray shaded regions are excluded by recent ATLAS SUSY squarks 
searches. Right column: contours of the production cross sections at $\sqrt{s}=14 \, \tev$ for $Z^\prime$ exotic decays.
}
\end{figure}

Two comments are in order here, regarding the discrepancies between collider searches 
discussed here for MSSM and G2HDM.
\begin{enumerate}

\item The signals of 2$ l +\mET$ and 2$ \tau +\mET$ in Spectrum-A$^\prime$ mainly come from Drell-Yan processes for both G2HDM and MSSM. The major difference between the two models is the distribution of the invariant mass of the final charged leptons. In G2HDM, the invariant mass distributions have a cut-off at $m_Z^\prime$, i.e., $ m_{l^{+}l^{-}}, \ m_{\tau^{+}\tau^{-}} < m_{Z'} $,
while $ m_{l^{+}l^{-}}$ and $m_{\tau^{+}\tau^{-}}$ are more evenly distributed in MSSM.
This is because the underlying Drell-Yan processes are mostly mediated by the on-shell $ Z' $ 
in G2HDM,
but by the off-shell SM $\gamma$ and $Z$ for MSSM, as indicated in Eqs.~\eqref{eq:slepton} and \eqref{eq:lH}.

\item 
If the on-shell $Z^\prime$ is highly boosted and the mass splitting between $Z^\prime$ and the new leptons is large, one will have two collinear outgoing new leptons
which result in two collinear SM leptons. In contrast, MSSM will not exhibit such a collinear behavior due to lack of $Z^\prime$, and so one can distinguish G2HDM from MSSM
via the event topology of dilepton plus missing transverse energy signals.

\end{enumerate}

\item

Spectrum-B$^\prime$: $ 2\,m_{ ( L^H,Q^H ) } < m_{Z^\prime}$ and 
$m_{\rm DM} = 50$ GeV. 

For the new quarks, they can always be pair produced dominantly by strong processes, like 
$q \bar q , gg  \to Q^H \overline{Q^H}$ via $s$-channel gluon exchange or $t$-channel heavy exotic
quark exchange.
The cross sections for the strong processes 
\begin{align}
p p\rightarrow  j^H\overline{j^H} \rightarrow 2j +\mET \,,\\
p p\rightarrow  b^H\overline{b^H} \rightarrow 2b +\mET \,,\\
p p\rightarrow  t^H\overline{t^H} \rightarrow 2t +\mET \,,
\end{align}
where $ j^H=(u^H, d^H, c^H, s^H) $ and $ j=(u, d, c, s) $
are computed.
On the other hand, processes involving intermediate squarks can lead to exactly the same final states, $2j +\mET$, $2b +\mET$ and $2t +\mET$.
Consequently, the LHC results of SUSY squark searches~\cite{ATLAS:2017cjl,Aaboud:2016nwl,ATLAS:2017kyf}
can be directly used for our case.

First, we choose the signal region \texttt{2j-1200}~\cite{ATLAS:2017cjl} for the $2j +\mET$ channel, which   
yields $\langle\epsilon\sigma\rangle^{95}_{\rm obs}\leq 3.6 $ fb  with 
$ \epsilon\sim 35\%$.
Second, for the $ 2b +\mET$ channel, the signal region \texttt{SRA250}~\cite{Aaboud:2016nwl} 
is chosen.
The resulting limit is $ \langle\epsilon A\sigma\rangle^{95}_{\rm obs}\leq 3.42 $ fb  with 
$ \epsilon A\sim 10\% $. 
Third, for the $ 2t +\mET$ channel, the signal region \texttt{SRA-T0}~\cite{ATLAS:2017kyf} 
is involved
and the corresponding limit is $ \langle\epsilon A\sigma\rangle^{95}_{\rm obs}\leq 0.40 $ fb with $ \epsilon A\sim 9\% $.
In the left column of Fig.~\ref{fig:SPB}, the red lines represent these cross sections, 
dominated by the strong interactions, at $\sqrt{s}=13\,\tev$ 
as functions of $ m_{Q^H} $, while the gray shaded regions are excluded 
by the recent ATLAS 13 TeV results.  
The cross section bounds can be translated into the new quark mass limits.
Clearly, from the left column of Fig.~\ref{fig:SPB}, 
we have $ m_{j^H}\gtrsim 1.4 \tev$, $ m_{b^H}\gtrsim 1 \tev$, and $ m_{t^H}\gtrsim 1.3\,\tev $.
Note that these mass bounds are independent of the $SU(2)_H$ coupling $g_H$
since the dominant cross sections here arise from pure QCD interactions.
 
On the other hand, if $ m_{Z^\prime} > 2 \, m_{Q^H} $, three subdominant but nevertheless 
important processes have to be included:
\begin{align}
\label{Zp2jmET}
p p\rightarrow Z' \rightarrow j^H\overline{j^H} \rightarrow 2j +\mET \,,\\
\label{Zp2bmET}
p p\rightarrow Z' \rightarrow b^H\overline{b^H} \rightarrow 2b +\mET \,,\\
\label{Zp2tmET}
p p\rightarrow Z' \rightarrow t^H\overline{t^H} \rightarrow 2t +\mET \, .
\end{align}
We display the corresponding cross section contours, similar to Fig.~\ref{fig:SPA} in Spectrum-A$^\prime$,
on the ($m_{Z'},m_{j^H}$), ($m_{Z'},m_{b^H}$), and ($m_{Z'},m_{t^H}$) planes respectively
in the right column of Fig.~\ref{fig:SPB}.  
The previous bounds on the new quark masses translated from the SUSY searches are also shown by the dashed lines. 
Because of the stringent mass limits which push the new quark mass scale beyond TeV,  
the production cross sections of the new quarks via the $Z^\prime$ exotic decays are kinematically suppressed and hence
constrained to be small: $\sigma/g^2_H \lesssim 0.01$ pb  as can be seen from the right column of Fig.~\ref{fig:SPB}.

\end{itemize}

\section{Future $W'$ Searches}
\label{sec:Wprime}

In the event that $Z'$ is discovered at HL-LHC via dijet or dilepton searches, one can ask
whether it comes from an additional $SU(2)$ gauge symmetry or simply from an extra $U(1)^\prime$.
In this section, we discuss how to look for the electrically neutral $W^{\prime (p,m)}$ whose existence
will help to pin down the $SU(2)_H$ as a potential underlying symmetry in nature.
In the rest of our analysis, we will switch back to Spectrum A and Spectrum B.

\subsection{$W^\prime$ in different $SU(2)$ models}

Before embarking on our detailed analysis, we should point out that there exist, 
of course, many other well-motivated models with neutral $W^\prime$
gauge bosons in addition to $Z^\prime$, 
such as THM~\cite{Chacko:2005pe}, 3-3-1 models~\cite{331}, {\it etc}. 
Certainly, any non-abelian gauge group commutes with the SM $SU(2)_L \times U(1)_Y$ gauge groups, 
naturally accommodates neutral $W^\prime$. 
Here we will not manage to compare the $W^{\prime (p,m)}$ of G2HDM with all models 
featuring neutral $W^\prime$.
Instead, we focus on collider signals of $W^{\prime (p,m)}$ in our model and contrast them with
LHT with T-parity -- a representative composite Higgs model 
which has neutral $Z^\prime$ and charged $W^\prime$~\cite{Cheng:2003ju}. 
Both $Z^\prime$ and $W^\prime$ in LHT can only be produced in pairs 
due to T-parity and have the same signals just like $W^{\prime (p,m)}$ in G2HDM.
In other words, LHT is chosen as an illustrative example to underscore
differences in the context of collider searches.  
Since $W^{\prime (p,m)}$ in G2HDM is always heavier than $Z^\prime$, 
it might not be easy to produce a pair of $W^{\prime (p,m)}$ (or even $Z^\prime$) at the LHC,
we will focus on the future 100 TeV proton-proton collider.
To identify $SU(2)_H$ unambiguously, 
the discovery of new heavy fermions $Q^H$ and $L^H$ as well as scalars 
like $\Delta_H$ and $\Phi_H$ will also be necessary on top of $W^{\prime (p,m)}$ and $Z^\prime$.

We note that LRSM~\cite{LRSM} has 
the right-handed charged $W_R^\pm$. It can be singly
produced and directly probed 
by dijet resonance~\cite{Dev:2015pga,Helo:2015ffa} or 
same-sign dilepton plus two jets ($l^\pm l^\pm jj$) 
searches~\cite{Keung:1983uu} depending on the right-handed neutrino mass,
whereas $W^{\prime (p,m)}$ in G2HDM must be pair produced.
Due to quite different properties between $W^{\prime (p,m)}$ and $W_R^\pm$ and
the stringent bound on $Z_R$: $m_{Z_R}> 3.2$ TeV~\cite{Lindner:2016lpp},
LRSM will not be considered here. 

The LHT model discussed here is based on the coset manifold $SU(5)/SO(5)$ 
which can be realized as a nonlinear sigma model~\cite{Cheng:2003ju}.
Two different  $SU(2) \times U(1)$ subgroups of  $SU(5)$ are gauged and are
broken down to the SM electroweak gauge group $SU(2)_L \times U(1)_Y$ 
at a scale $f_T$, which is higher than but not too far away from the electroweak scale 
so as to provide a possible solution to the fine-tuning problem.
In the LHT model, all particles are divided into two classes based on the T-parity 
(denoted by $\mathcal{P}_T$ hereafter), which corresponds to the symmetry 
under the exchange of the two $SU(2) \times U(1)$ subgroups.
As a result, combinations of different fields 
of the two subgroups can be formed as having eigenvalue $+ 1$ or $-1$
of $\mathcal{P}_T$. 
The lightest T-odd particle is $A_H$, a spin 1 particle, which is ensured to be stable 
and hence can be a DM candidate. Novel collider signatures like monojet and dijet plus
missing transverse energy of $A_H$ in LHT was studied in~\cite{Chen:2006ie}.
All of the exotic particles have their masses proportional to $f_T$ 
since the masses are induced from the collective symmetry breaking at the scale $f_T$. 
Furthermore, the exotic fermions couple to T-odd combinations of the gauge bosons 
of $(SU(2) \times U(1))^2$. 
Three  of the combinations comprise a non-abelian group $SU(2)_T$ which is 
broken at the scale $f_T$.
In the end, the exotic T-odd $SU(2)_T$ gauge bosons should  couple to one T-even 
and one T-odd particles so as to conserve the T-parity. 
We will use $Z_H$ and $W_H^{\pm}$ to denote the $SU(2)_T$ gauge bosons in the LHT, 
as opposed to $Z'$ and $W^{\prime (p,m) }$ in G2HDM.
The quantum numbers of additional $SU(2)_L$ singlet fermions in the G2HDM and LHT are summarized in Table~\ref{Tab:assignments}. 
Note that the superscript $H$ is specifically used to indicate the G2HDM exotic fermions,
while the subscript $H$ denotes the LHT new fermions.

\begin{table}
\begin{tabular}{|c||c|c|c|c|c|c|c|c|c|c|c|c|c|}
\hline
 &\multicolumn{8}{|c|}{G2HDM}  &\multicolumn{5}{|c|}{LHT}\\
\hline
 &$U_R$ &$D_R$ & $u^H_L$ & $d^H_L $ & $N_R$ &$E_R$ & $\nu^H_L$ & $e^H_L $ & $q_H$ &$t_H$ &$d_H$ &$l_H$ &$e_H$\\
\hline
$SU(3)_C$ &3 &3 &3 &3 &1 &1 &1 &1&3 &3 &3 &1 &1\\
\hline
$SU(2)_H$ &2 &2 &1 &1 &2 &2 &1 &1&$\raisebox{1mm}{\cancel{\makebox[2.5mm]{}}}$ &$\raisebox{1mm}{\cancel{\makebox[2.5mm]{}}}$ &$\raisebox{1mm}{\cancel{\makebox[2.5mm]{}}}$ &$\raisebox{1mm}{\cancel{\makebox[2.5mm]{}}}$ &$\raisebox{1mm}{\cancel{\makebox[2.5mm]{}}}$\\
\hline\hline
$SU(2)_T$ &$\raisebox{1mm}{\cancel{\makebox[2.5mm]{}}}$ &$\raisebox{1mm}{\cancel{\makebox[2.5mm]{}}}$ &$\raisebox{1mm}{\cancel{\makebox[2.5mm]{}}}$ &$\raisebox{1mm}{\cancel{\makebox[2.5mm]{}}}$ &$\raisebox{1mm}{\cancel{\makebox[2.5mm]{}}}$ &$\raisebox{1mm}{\cancel{\makebox[2.5mm]{}}}$ &$\raisebox{1mm}{\cancel{\makebox[2.5mm]{}}}$ &$\raisebox{1mm}{\cancel{\makebox[2.5mm]{}}}$ &2 &1 &1 &2 &1\\
\hline
$\mathcal{P}_T$ &$\raisebox{1mm}{\cancel{\makebox[2.5mm]{}}}$ &$\raisebox{1mm}{\cancel{\makebox[2.5mm]{}}}$ &$\raisebox{1mm}{\cancel{\makebox[2.5mm]{}}}$ &$\raisebox{1mm}{\cancel{\makebox[2.5mm]{}}}$ &$\raisebox{1mm}{\cancel{\makebox[2.5mm]{}}}$ &$\raisebox{1mm}{\cancel{\makebox[2.5mm]{}}}$ &$\raisebox{1mm}{\cancel{\makebox[2.5mm]{}}}$ &$\raisebox{1mm}{\cancel{\makebox[2.5mm]{}}}$ &$-1$ &$-1$ &$-1$ &$-1$ &$-1$\\
\hline
\end{tabular}
\caption{Comparison of quantum numbers of the heavy $SU(2)_L$ singlet fermion fields in G2HDM and LHT. In case of the absence of the symmetries in the models, we put a slash in the cells. $\mathcal{P}_T$ is the T-parity in LHT.
}
\label{Tab:assignments}
\end{table}

\subsection{Two search channels: $ 2l +\mET $ and $ 4l +\mET $}

In G2HDM, the $W^{\prime (p,m)}$ boson pair are  produced via
the $Z^\prime$ and $Q^H$ exchange, while  
the pair productions of $W^+_H W^-_H$ and $Z_H Z_H$ are through the $\gamma$, $Z$, and $q_H$ exchange in LHT.
We will focus on leptonic decay channels for these gauge boson pairs because of the low QCD background
as in the $Z^\prime$ resonance searches.
The final states of two and four leptons plus the missing transverse energy, $ 2l +\cancel{\it{E}}_{T} $
and $ 4l +\cancel{\it{E}}_{T}$ respectively, will be investigated.

Take the $W^{\prime (p,m)}$ pair in G2HDM as an example. 
The $ 2l +\cancel{\it{E}}_{T}$ channel comes from the prompt decay of one of the two gauge bosons into one heavy charged lepton plus one SM lepton, while the other boson into one heavy neutrino plus one light neutrino. 
Each of the two resulting heavy fermions then
decays into the DM particle $H_2^0$ plus the corresponding light SM fermion through the Yukawa couplings from~\eqref{eq:Yuk_SM}. The neutrinos and DM particles in the final state 
will escape from the detector and manifest as $\cancel{\it{E}}_{T}$. 
In Table~\ref{tab:signature}, we summarize the decay chains of the gauge boson pairs into $ 2l +\cancel{\it{E}}_{T} $
and $ 4l +\cancel{\it{E}}_{T}$. In the last column labeled by Signal, the particles inside the curly brackets manifest as $\cancel{\it{E}}_{T}$.

\begin{table}[htp!]
\begin{tabular}{|c|c|c|c|c|}
\hline
\hline
 Model & Production & Prompt Decay & Final State & Signal  \\
\hline
\hline
      \multicolumn{5}{|c|}{$ l^{+}l^{-} +\cancel{\it{E}}_{T} $} \\
\hline
\hline
G2HDM & $p p\rightarrow W'^{p}W'^{m}$ & $(\overline{l}l^{H})(\overline{\nu^{H}}\nu)$  + c.c. &
 $(\overline{l} l H_2^{0})(\overline{\nu}H_2^{0*}\nu)$ & $l^+l^- +\lbrace\nu\overline{\nu}H_2^{0}H_2^{0*}\rbrace$ \\
\hline
LHT & $p p\rightarrow W^{+}_{H} W^{-}_{H}$ & $(l^{+}_{H}\nu)(\overline{\nu_{H}} l^{-})$  + c.c. &
 $(l^{+}A_{H}\nu)(\overline{\nu}A_{H}l^{-})$ & $l^{+}l^{-} +\lbrace \nu\overline{\nu} A_{H}A_{H}\rbrace$ \\
\hline
LHT & $p p\rightarrow Z_{H} Z_{H}$ & $(l^{\pm}_{H}l^{\mp})(\overline{\nu_{H}}\nu)$  &
 $(l^{\pm}A_{H}l^{\mp})(\overline{\nu}A_{H}\nu)$ & $l^{+}l^{-} +\lbrace\nu\overline{\nu}A_{H}A_{H}\rbrace$ \\
\hline
\hline
      \multicolumn{5}{|c|}{$ l^{+}l^{-}l^{+}l^{-}+\cancel{\it{E}}_{T} $} \\
\hline
\hline
G2HDM & $p p\rightarrow W'^{p}W'^{m}$ & $(\overline{l}l^{H})(l\overline{l^{H}})$  &
 $(\overline{l}lH_2^{0})(l\overline{l}H_2^{0\ast})$ & $l^{+}l^{-}l^{+}l^{-} +\lbrace H_2^{0}H_2^{0\ast}\rbrace$ \\
\hline
LHT & $p p\rightarrow Z_{H} Z_{H}$ & $(l^{\pm}_{H}l^{\mp})(l^{\pm}_{H}l^{\mp})$  &
 $(l^{\pm}A_{H}l^{\mp})(l^{\pm}A_{H}l^{\mp})$ & $l^{+}l^{-}l^{+}l^{-} +\lbrace A_{H}A_{H}\rbrace$ \\
\hline
\hline
\end{tabular}
\caption{List of the production and leptonic decay channels for the exotic gauge bosons in G2HDM and LHT.
}
\label{tab:signature}
\end{table}

\subsection{The quantitative study: cross sections}

\begin{figure}[th!]
\centering
\includegraphics[width=3.in, height=3.in]{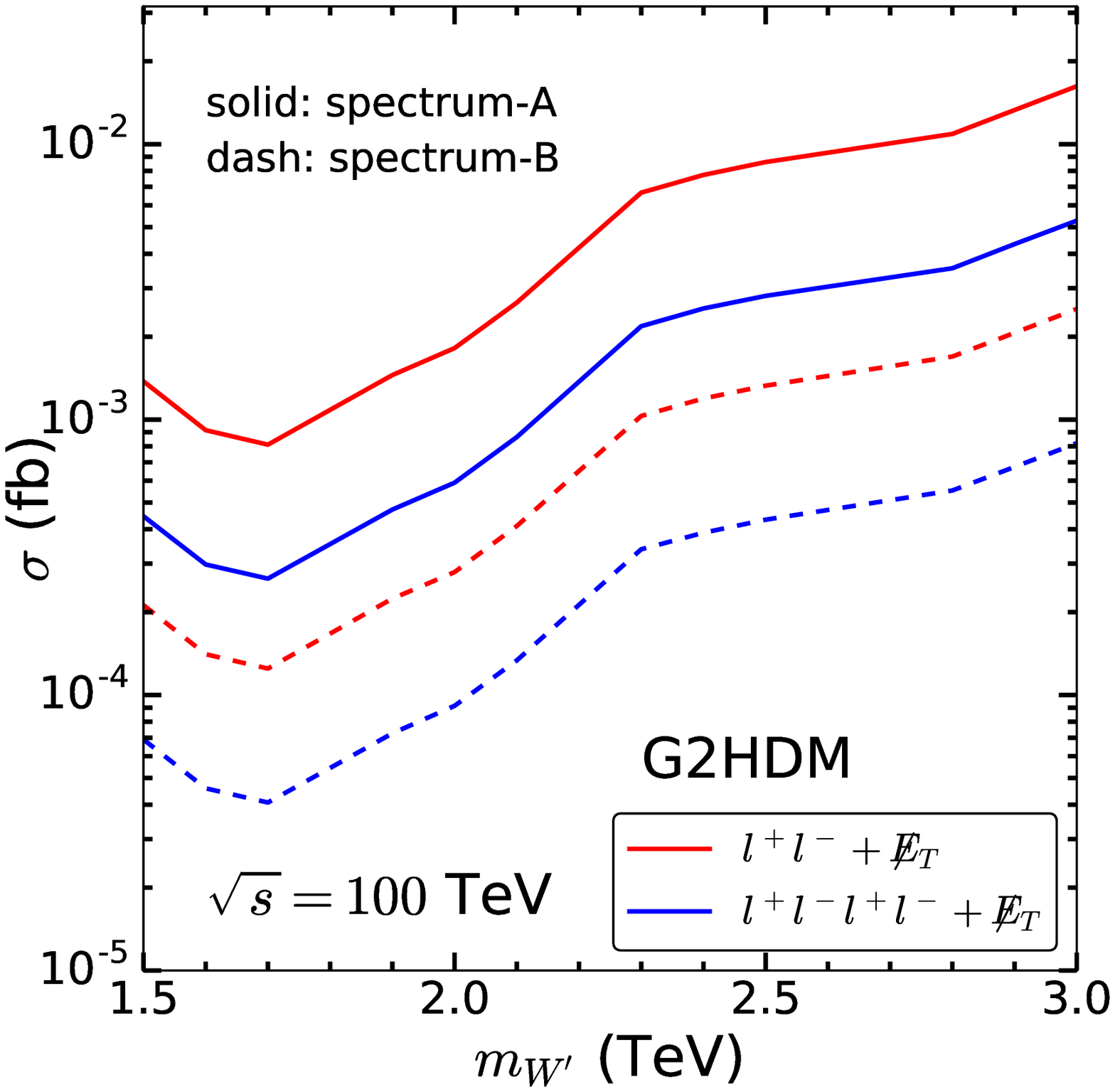}
\includegraphics[width=3.in, height=3.in]{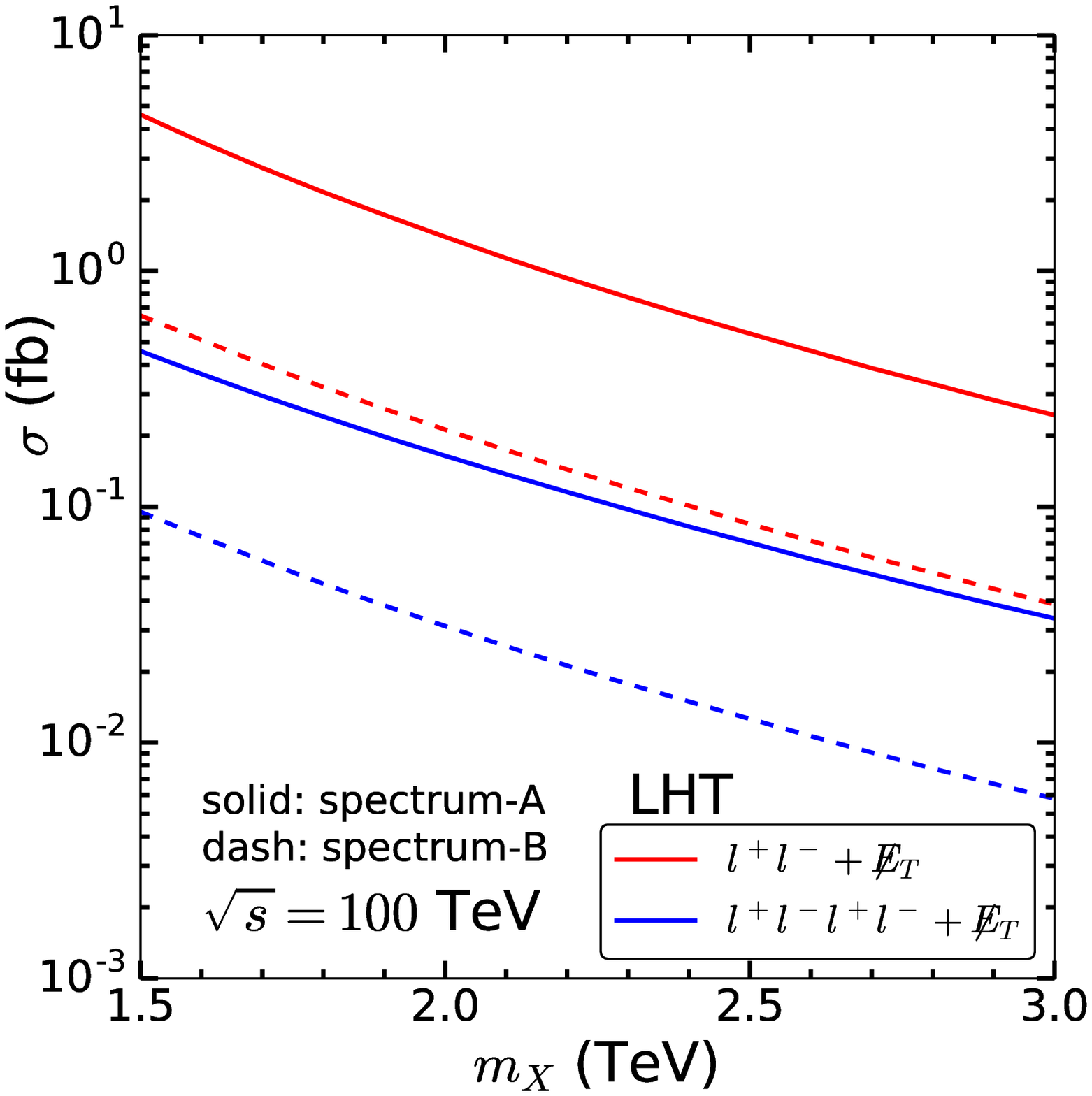}
\caption{\small \label{Fig:Xsec}
Left Panel: Cross section $ \sigma $ (fb) versus the mass $ m_{W^{\prime (p,m)}}$ (TeV) for $ 2l +\mET $ and $ 4l +\mET $ channels. 
The red/blue solid (dashed) lines are the cross sections of $ 2l +\mET $ and $ 4l +\mET $ channels computed at $\sqrt{s}=100\,\tev$ predicted by Spectrum-A (Spectrum-B)
in G2HDM, respectively. 
Right Panel: The LHT plot similar to the left panel. The $SU(2)_T$ gauge coupling is fixed to be the SM $SU(2)_L$ coupling, whereas
$g_H$ in G2HDM is set to be the maximally allowed value by the $Z'$ dilepton searches, presented in Fig.~\ref{fig:zp_con}.
}\end{figure}

In Fig.~\ref{Fig:Xsec}, we show the cross sections as functions of the corresponding gauge boson mass for the channels $ 2l +\cancel{\it{E}}_{T}$ and $4l + \cancel{\it{E}}_{T} $~($ l= e, \mu $) with $\sqrt{s}=100$ TeV in both G2HDM~(left panel) and LHT~(right panel).
The solid~(dashed) line corresponds to Spectrum-A~(Spectrum-B), while the red~(blue) line refers to the channel $ 2l +\cancel{\it{E}}_{T}$~($4l + \cancel{\it{E}}_{T}$).
Here we apply the same Spectrum A and Spectrum B for the gauge bosons, heavy exotic fermions and DM in G2HDM to the corresponding particles in LHT.

For G2HDM, the upper bound from the $Z^\prime$ resonance searches on the gauge coupling $g_H$ from Fig.~\ref{fig:zp_con} is used.
In other words, the region above the line in each case is excluded.
That is the reason why the cross sections, which scale as $g^4_H$, increase 
when $m_{W^\prime}$~(and also $m_{Z^\prime}$) becomes larger, 
since the bound on $g_H$ becomes less stringent.
By contrast, in LHT the gauge coupling is set equal to the SM electroweak coupling and thus the cross sections decrease as
the gauge boson mass increases.

We should point out that in G2HDM the process $p p\rightarrow Z' \rightarrow \overline{l^H} l^H$ is actually the 
dominant contribution to the final state 2$ l +\mET$. 
If $Z^\prime$ is discovered in the resonance searches, 
as assumed here, one should be able to infer the precise values of 
$g_H$ and $m_{Z^\prime}$.
Therefore, the dominant $Z^\prime$ contribution can be subtracted from the data so that one can study the contributions from $p p\rightarrow W'^p W'^m$ alone.
Alternatively, as we shall see later one can also resort to the $ 4l +\mET$ final state for  $W^{\prime (p,m)}$ 
searches. It is a relatively clean channel and is free from $p p\rightarrow Z'$ pollution.
The corresponding cross section is only 
three times smaller than that of the $ 2l +\cancel{\it{E}}_{T} $ channel.
Finally, the $Z'$ pair production $p p\rightarrow Z^\prime Z^\prime$ 
will also produce 
the $ 2l +\cancel{\it{E}}_{T} $ and $ 4l +\cancel{\it{E}}_{T} $ signals.
The contributions, however, are at least two orders of magnitude smaller than those from $p p\rightarrow W'^p W'^m$,
and therefore will be neglected in our analysis.

In LHT, the cross section of $ p p\rightarrow W^{+}_{H} W^{-}_{H} $ 
is about one order of magnitude larger than $ p p\rightarrow Z_{H} Z_{H} $ 
at $ \sqrt{s}=100 $ TeV. 
That is because $W^{+}_{H} W^{-}_{H} $ is produced dominantly by the $s$-channel $ \gamma$ and $Z$ exchange,
which is larger than the main contribution from the $t$-channel $q_H$ exchange to the $Z_{H} Z_{H} $ production.
Consequently, the cross section for $2l +\mET $, mostly from the $W^{+}_{H} W^{-}_{H}$ channel, 
is almost one order of magnitude larger 
than that of $ 4l +\mET$, which arises only from $Z_{H} Z_{H} $ channel.
As seen from Fig.~\ref{Fig:Xsec}, however, the cross sections for both 
$2l +\mET $ and $ 4l +\mET $ are of the same order in G2HDM.
On the other hand, the cross sections for both of these channels in LHT are
roughly $1\sim 2$ orders of magnitude larger than those in G2HDM, depending on the 
gauge boson mass.
Thus, one can in general distinguish the two models just by measuring the total 
cross sections of the two channels.  

\subsection{The qualitative study: the kinematical distributions}

In this section, we will further investigate the difference 
between G2HDM and LHT gauge boson decays in terms of three different normalized kinematical distributions.
In principle, one should be able to distinguish the electrically charged $W^{\pm}_H$ from the neutral $W^{\prime (p,m)}$
by the total charge of the corresponding decay products once they are 
produced singly.
Both $W^{\prime (p,m)}$ and $W^\pm_H$, however, have to be pair-produced 
because of the $SU(2)_H$ and $SU(2)_T$ symmetry respectively which lead to
the same total charge of the final states.    
As a consequence, the kinematical distributions of $W^{\prime (p,m)}$ and 
$W^\pm_H$ decays are not only interesting but also important to study for further information, 
even though the production cross sections, as shown in previous Section, 
are in general much larger in LHT than in G2HDM.

\begin{figure}[th!]
\centering
\includegraphics[width=2.65in, height=2.65in]{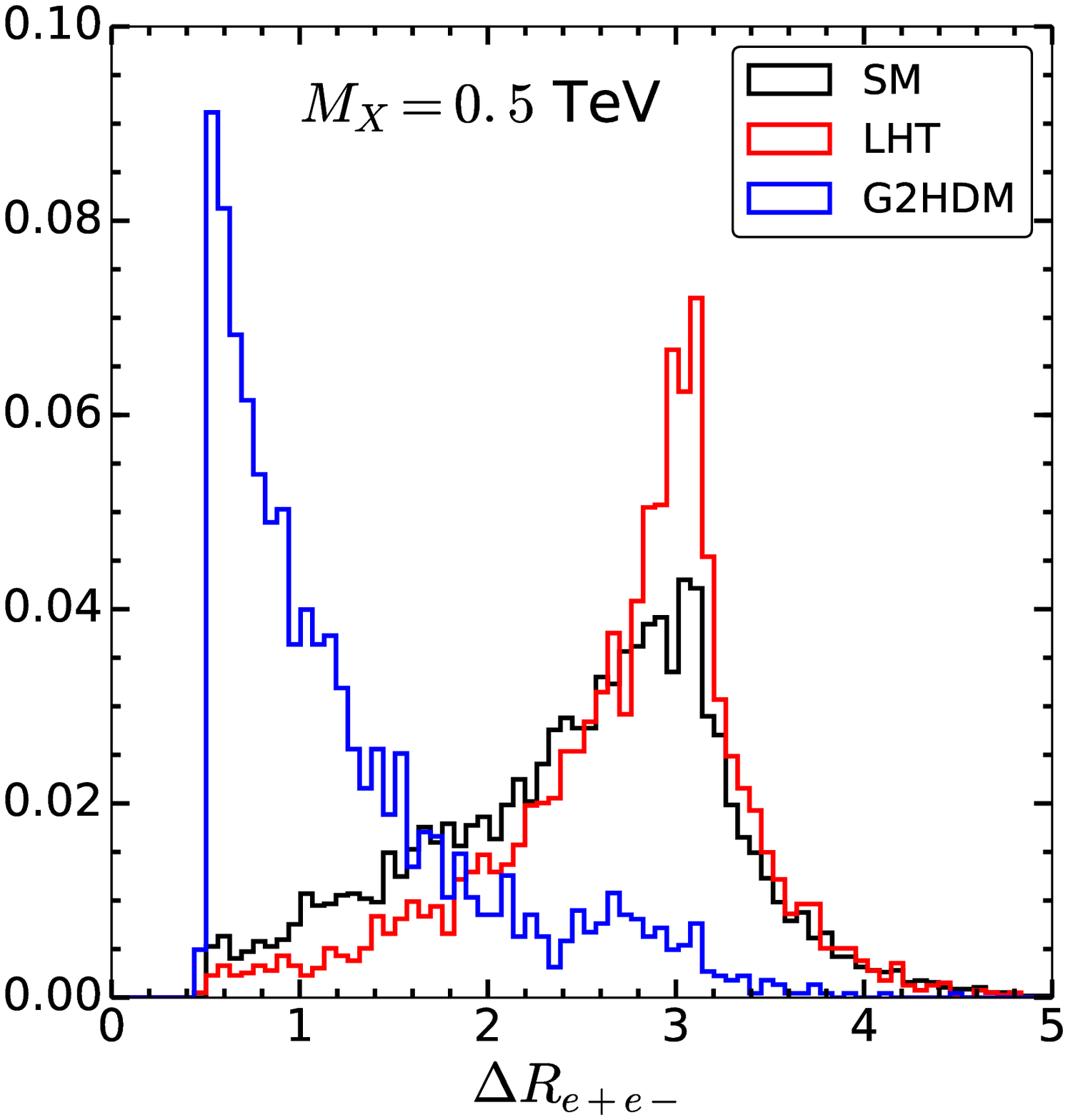}
\includegraphics[width=2.65in, height=2.65in]{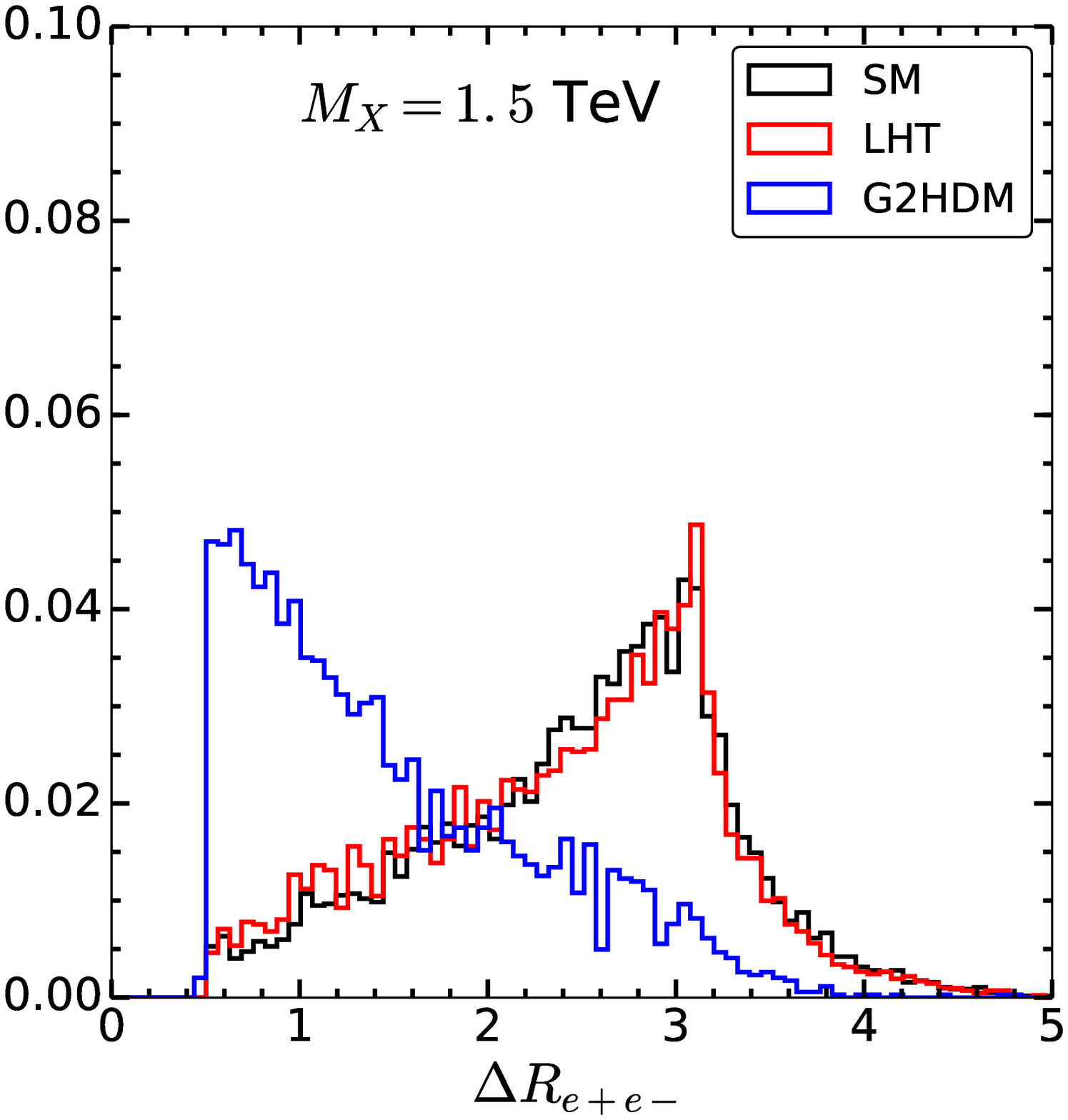}
\includegraphics[width=2.65in, height=2.65in]{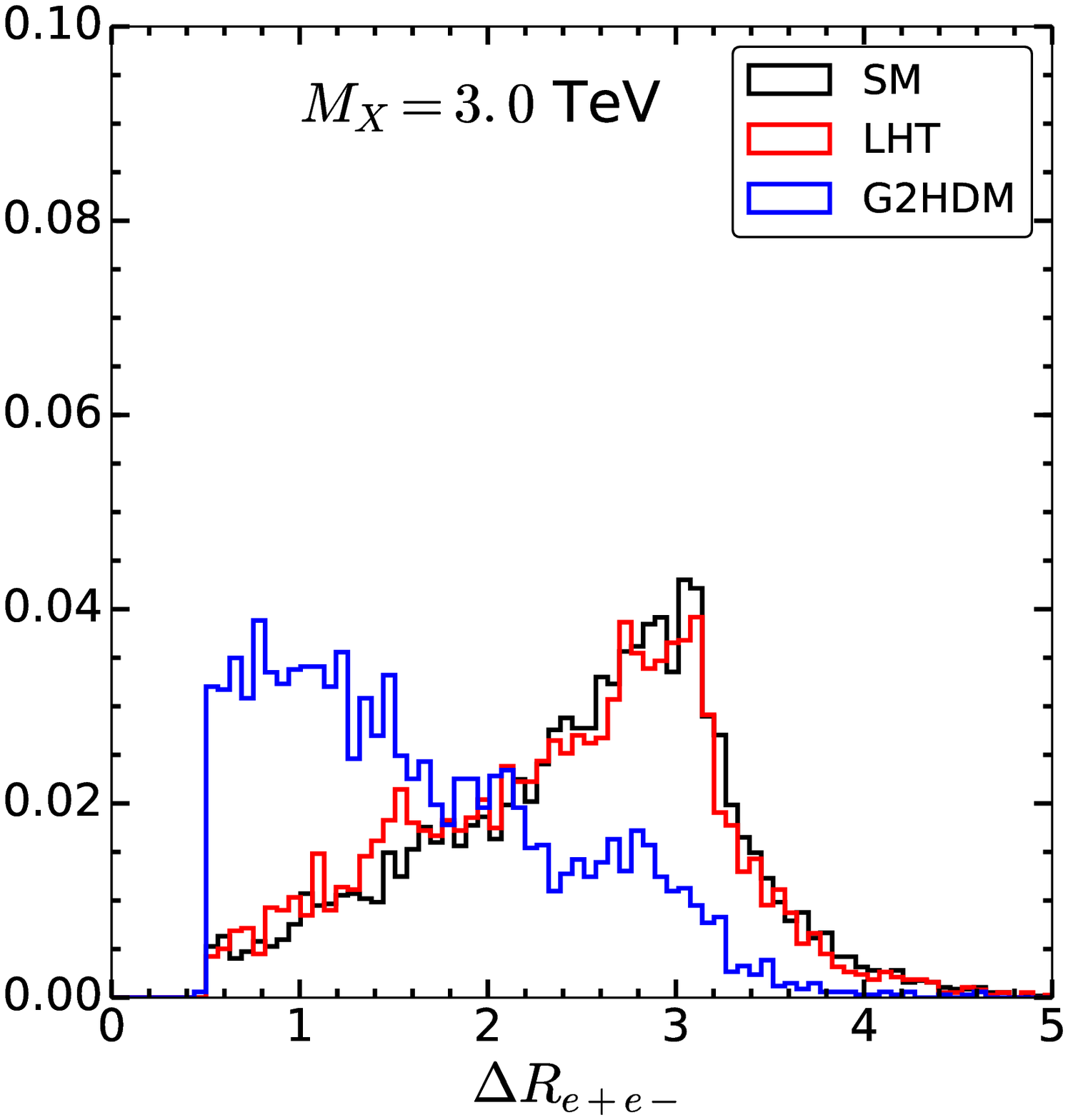}
\includegraphics[width=2.65in, height=2.65in]{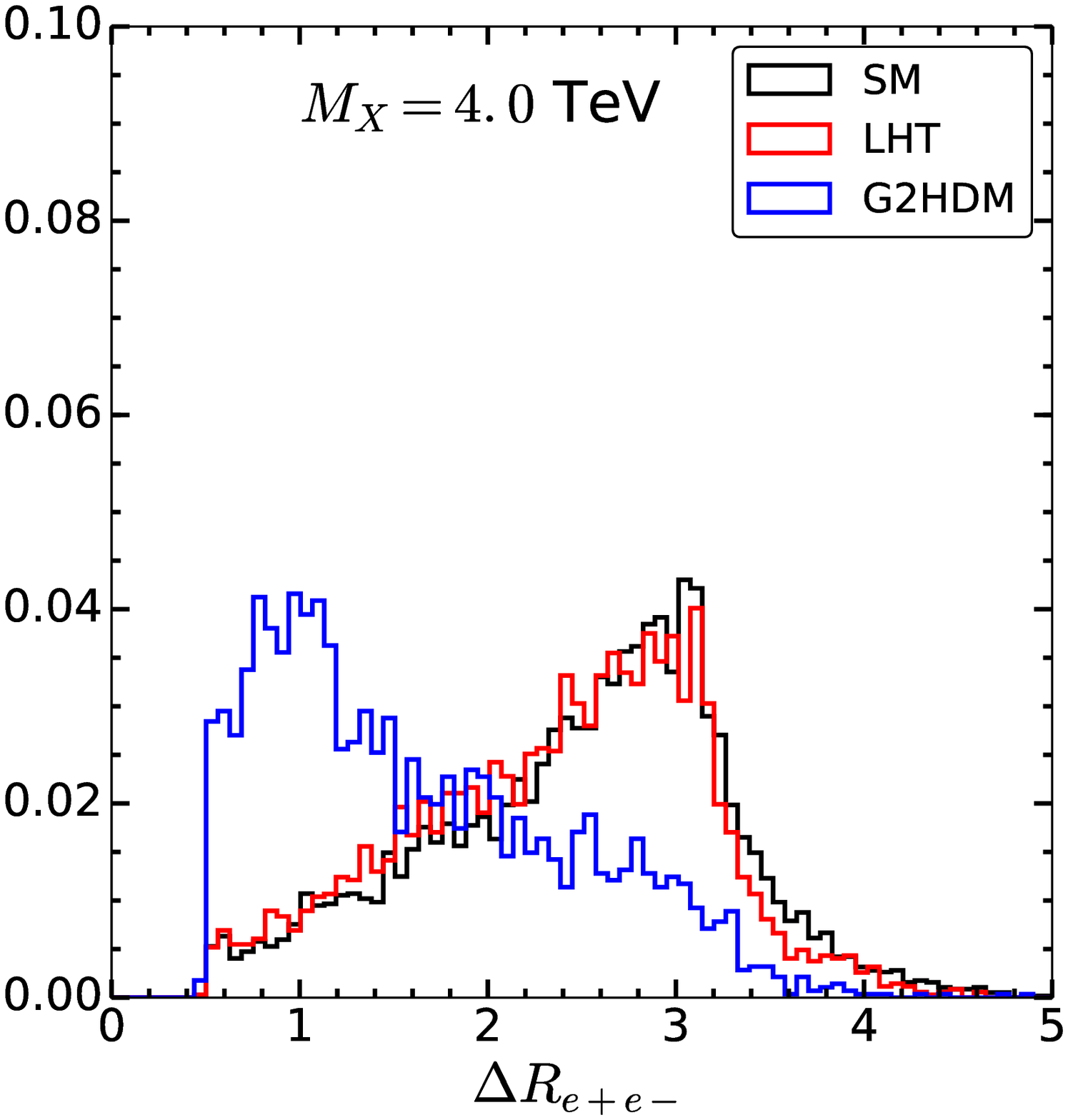}
\caption{\small \label{Fig:dR}
Comparison of the normalized distributions of the spatial separation $ \bigtriangleup R_{e^{+}e^{-}} $ in
G2HDM (blue), LHT (red), and SM (black) computed at $ \sqrt{s}= $ 100 TeV 
with $ m_{X}=0.5\,\tev$ (top-left), $ 1.5\,\tev$ (top-right), $ 3.0\,\tev$ (bottom-left), and $ 4.0\,\tev$ (bottom-right).
}
\end{figure}

\begin{figure}[th!]
\centering
\includegraphics[width=2.65in, height=2.65in]{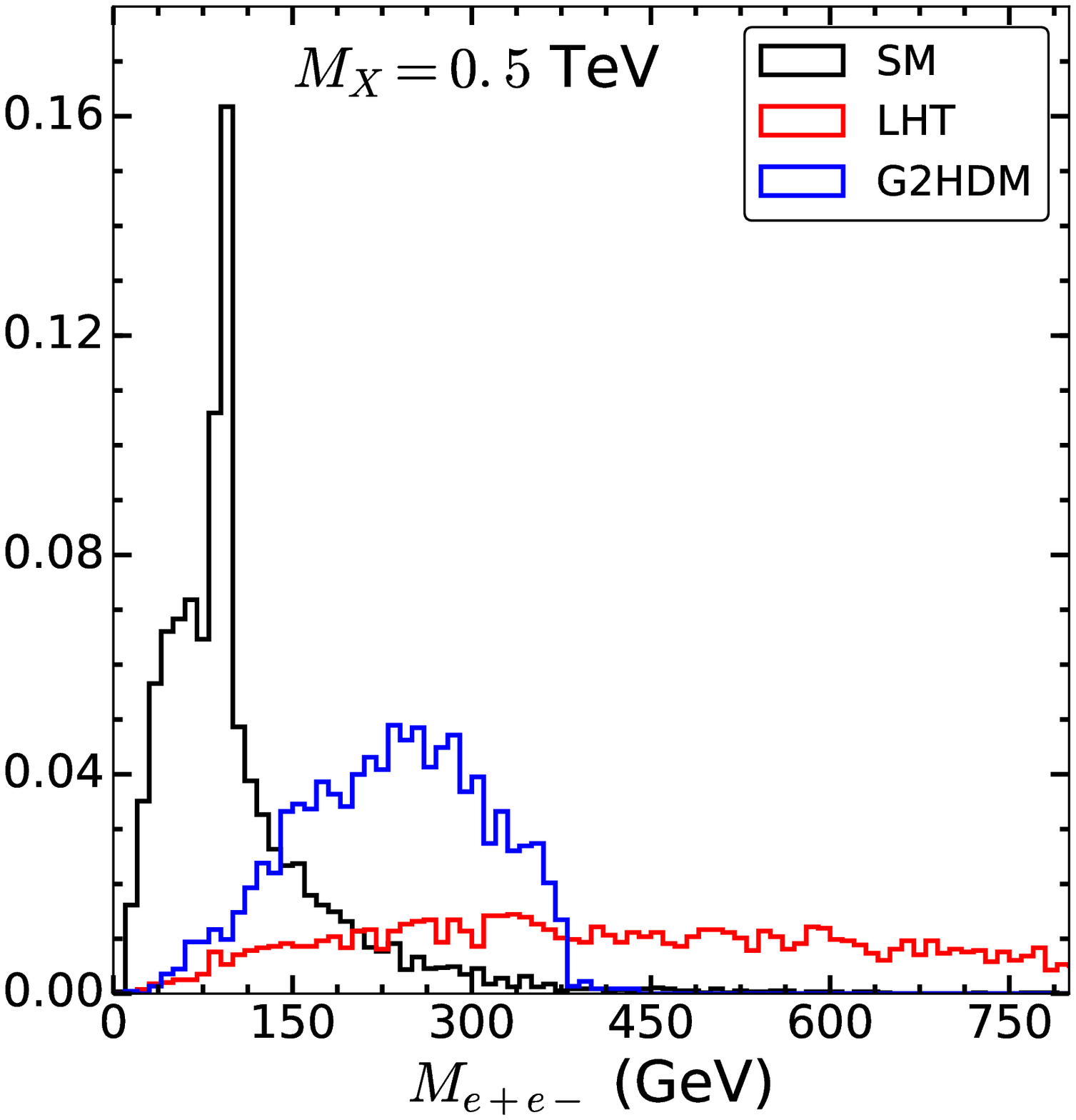}
\includegraphics[width=2.65in, height=2.65in]{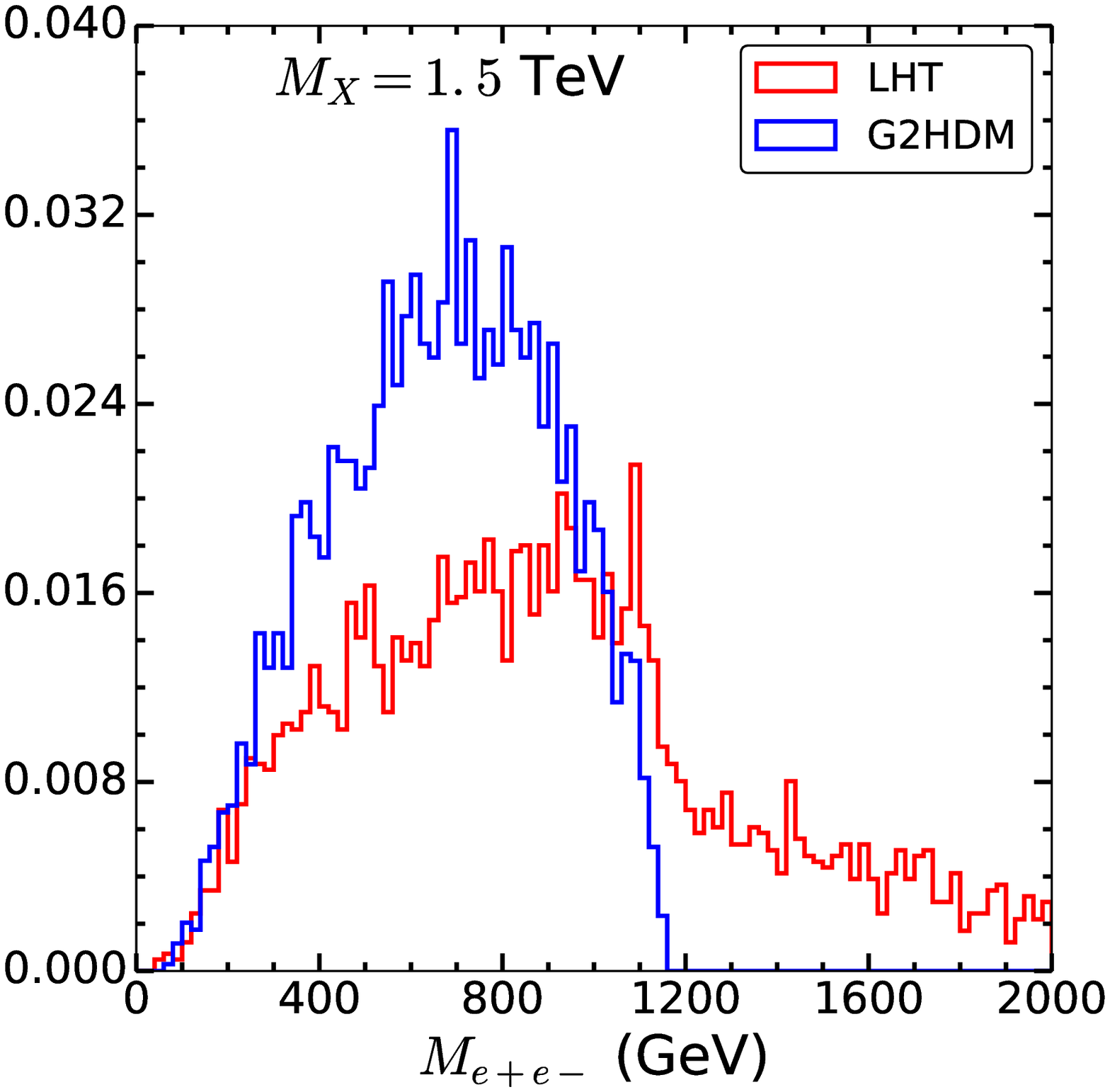}
\includegraphics[width=2.65in, height=2.65in]{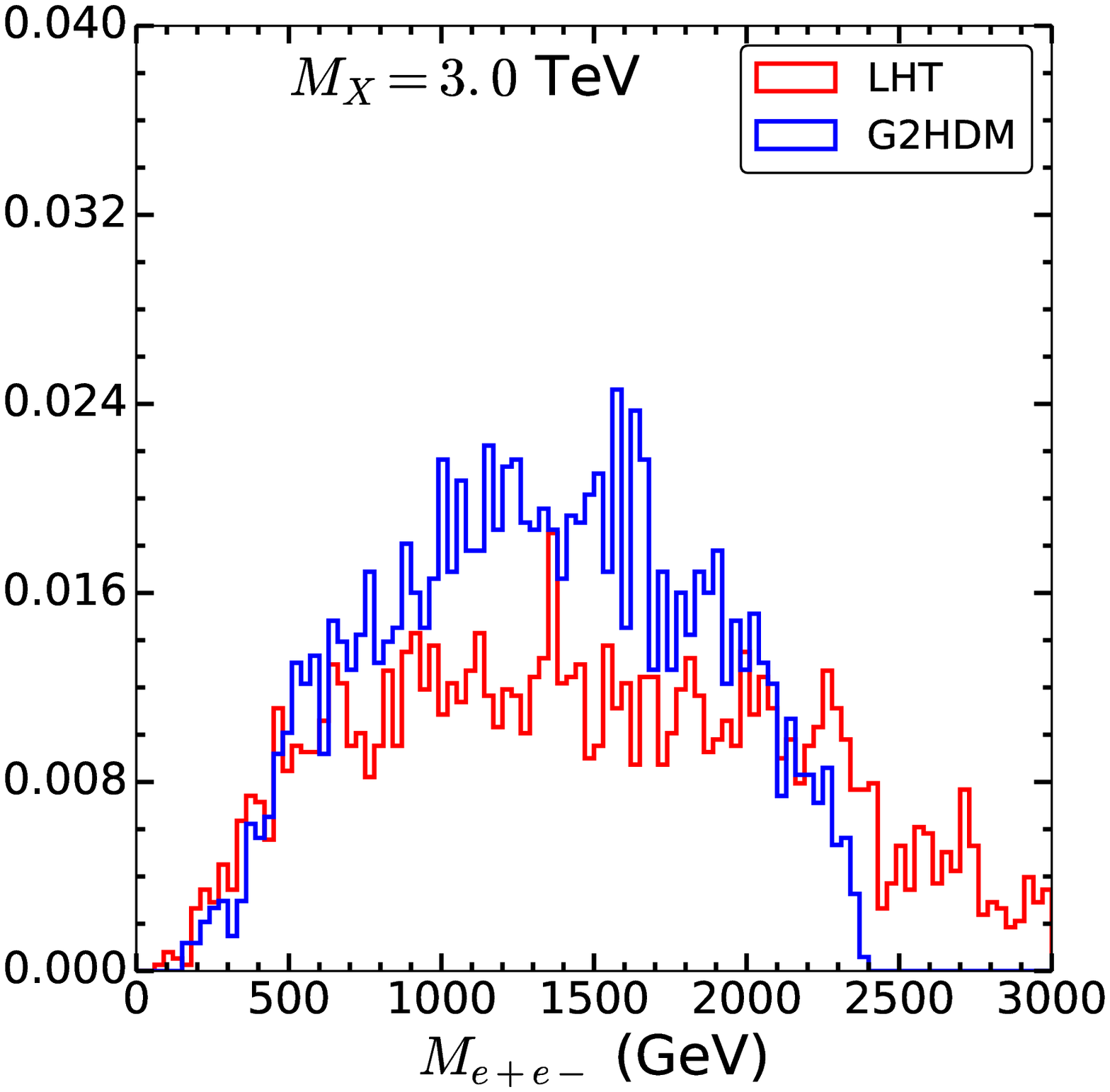}
\includegraphics[width=2.65in, height=2.65in]{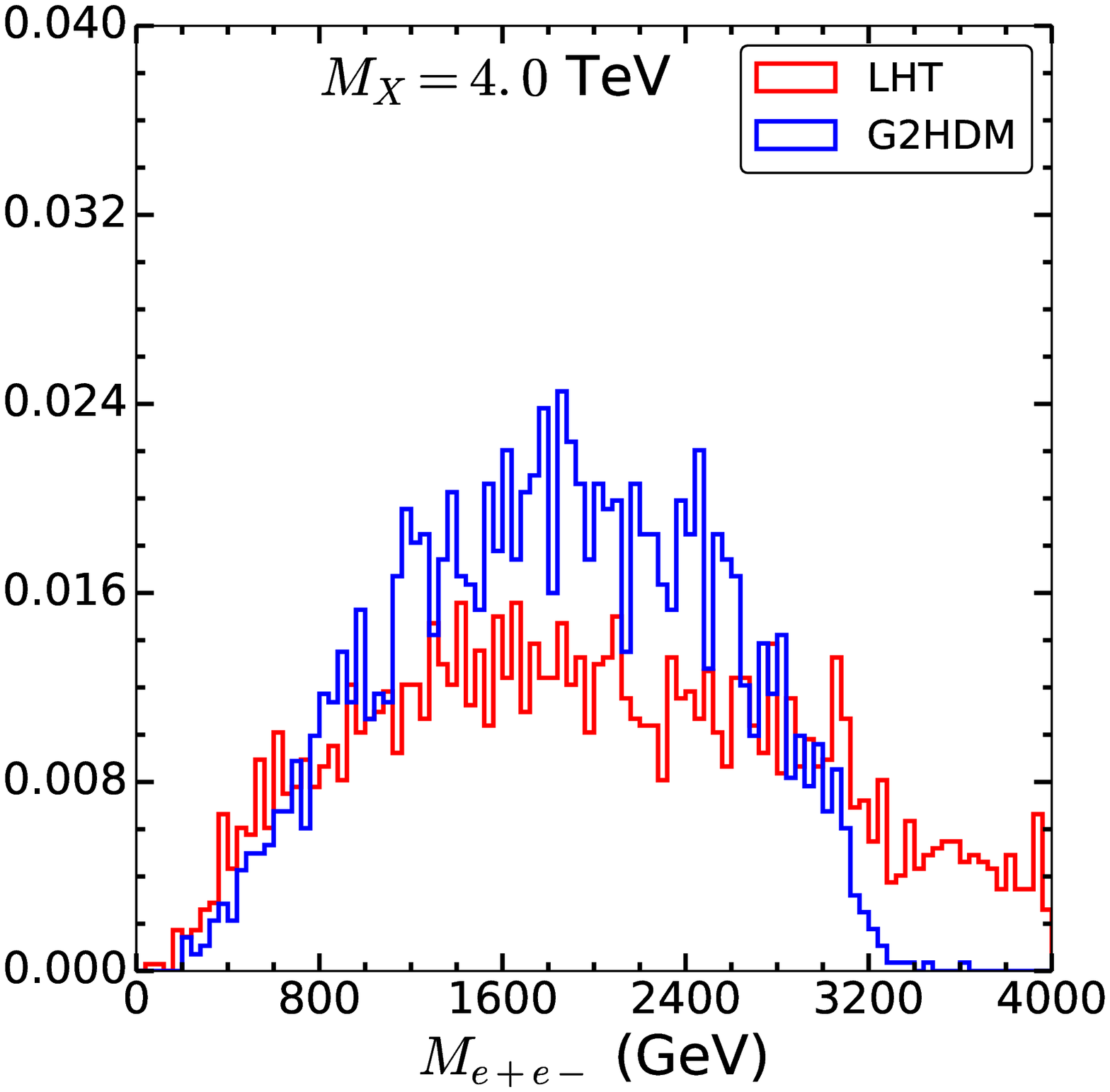}
\caption{\small \label{Fig:IM}
Comparison of the normalized distributions of the invariant mass $ M_{e^{+}e^{-}} $ in
G2HDM (blue), LHT (red), and SM (black) computed at $ \sqrt{s}= $ 100 TeV 
with $ m_{X}=0.5\, \tev$ (top-left), $ 1.5\,\tev$ (top-right), $ 3.0\,\tev$ (bottom-left), and $ 4.0\,\tev$ (bottom-right).
}
\end{figure}

\begin{figure}[th!]
\centering
\includegraphics[width=2.65in, height=2.65in]{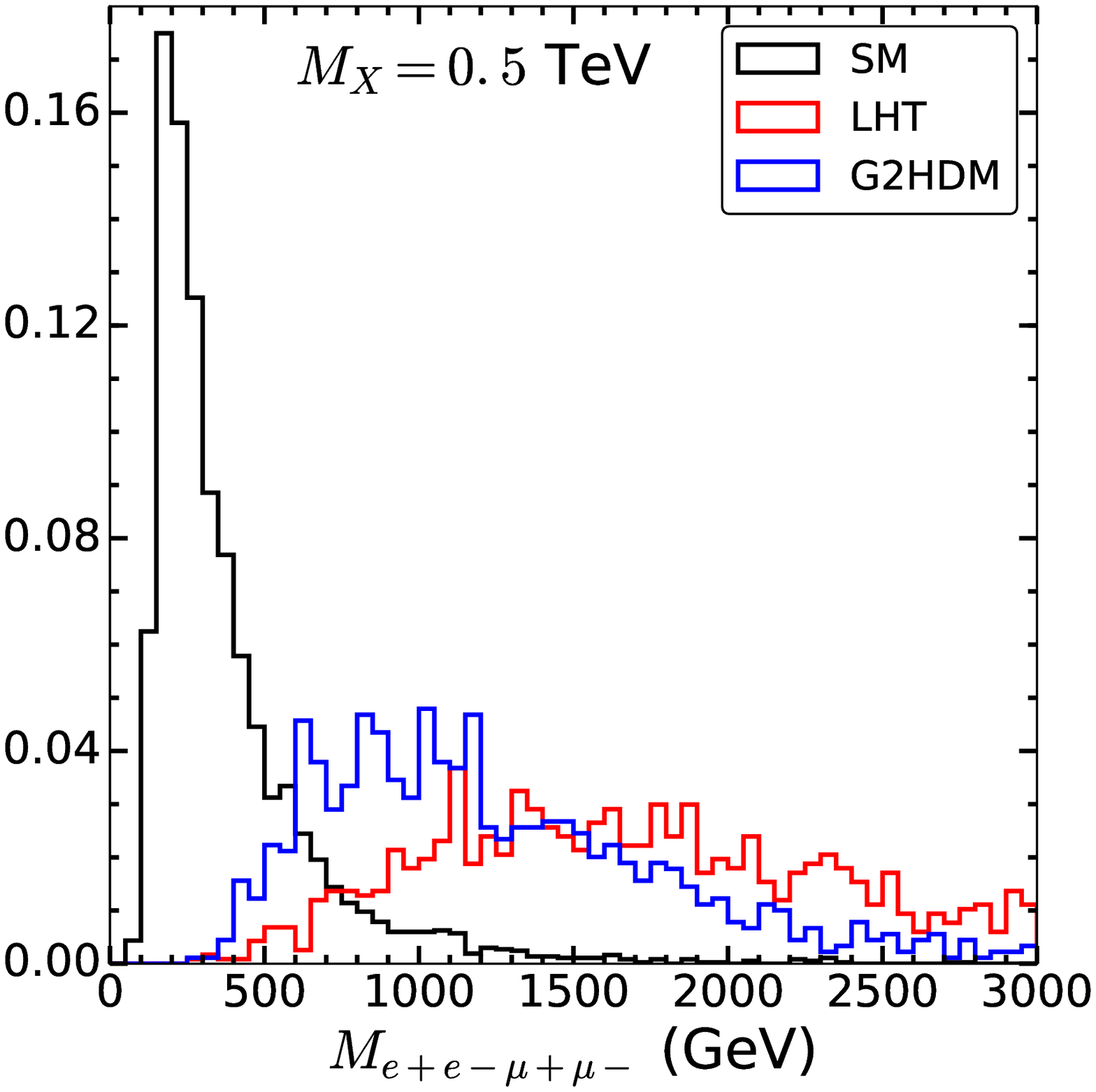}
\includegraphics[width=2.65in, height=2.65in]{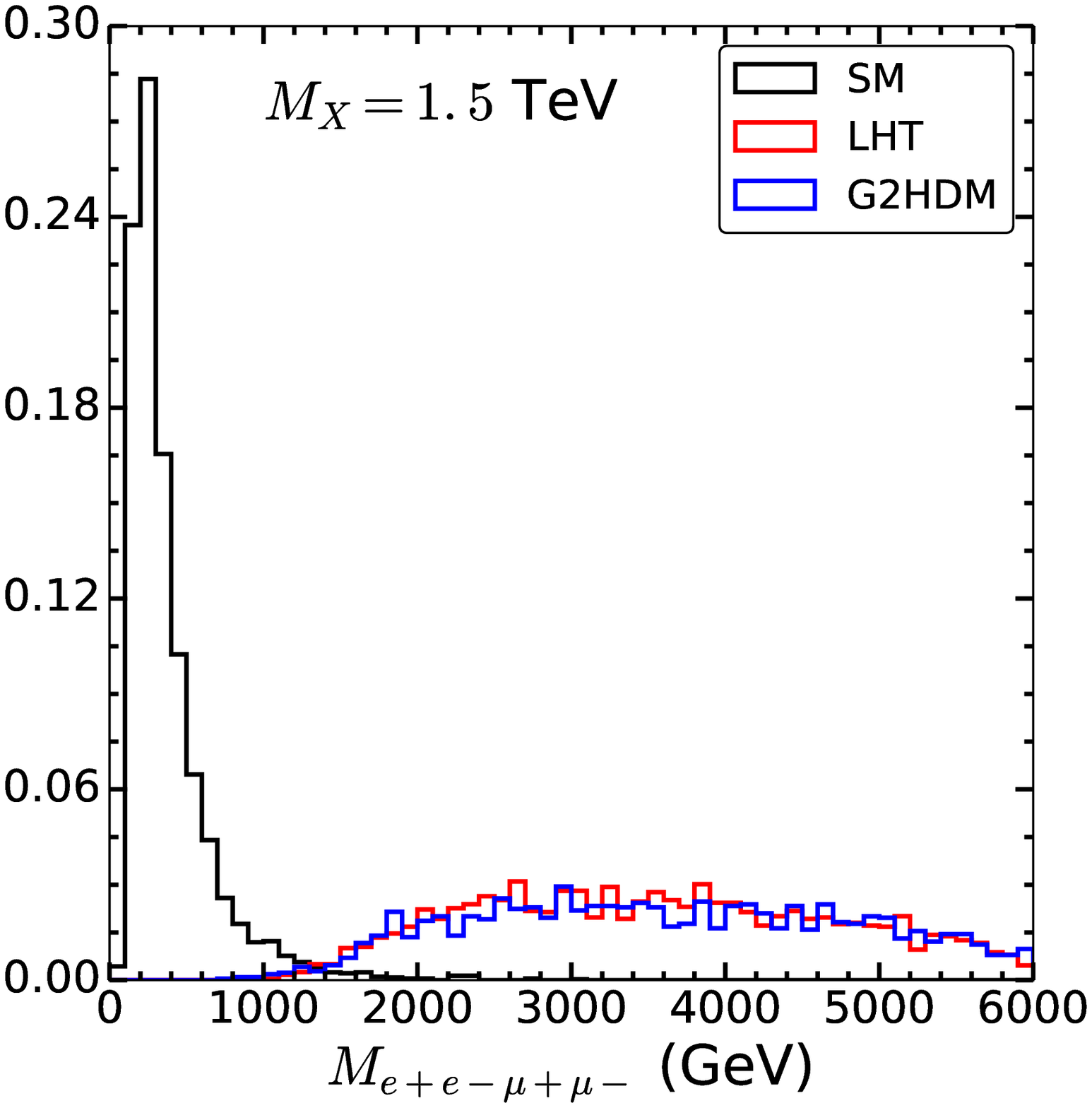}
\includegraphics[width=2.65in, height=2.65in]{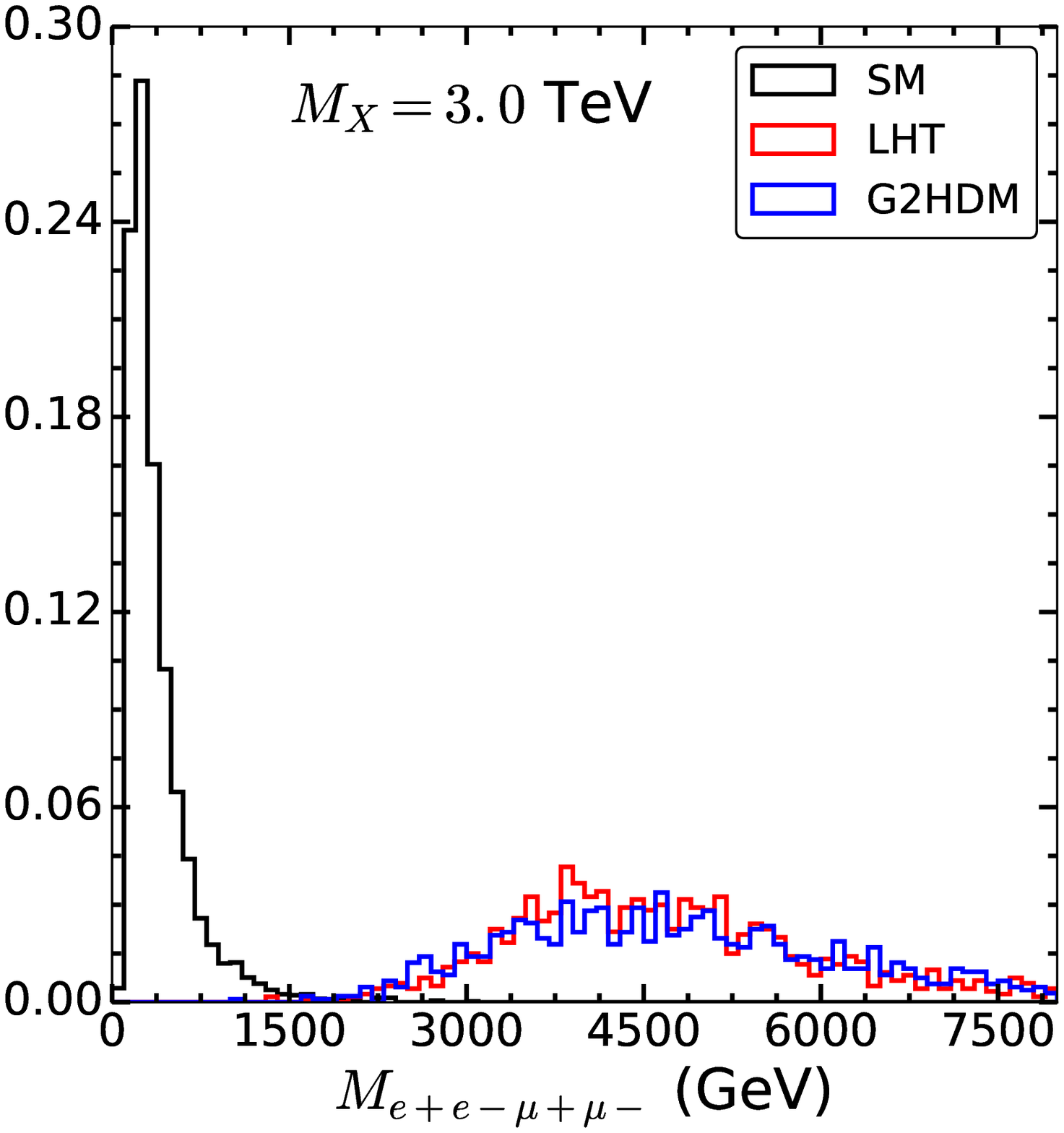}
\includegraphics[width=2.65in, height=2.65in]{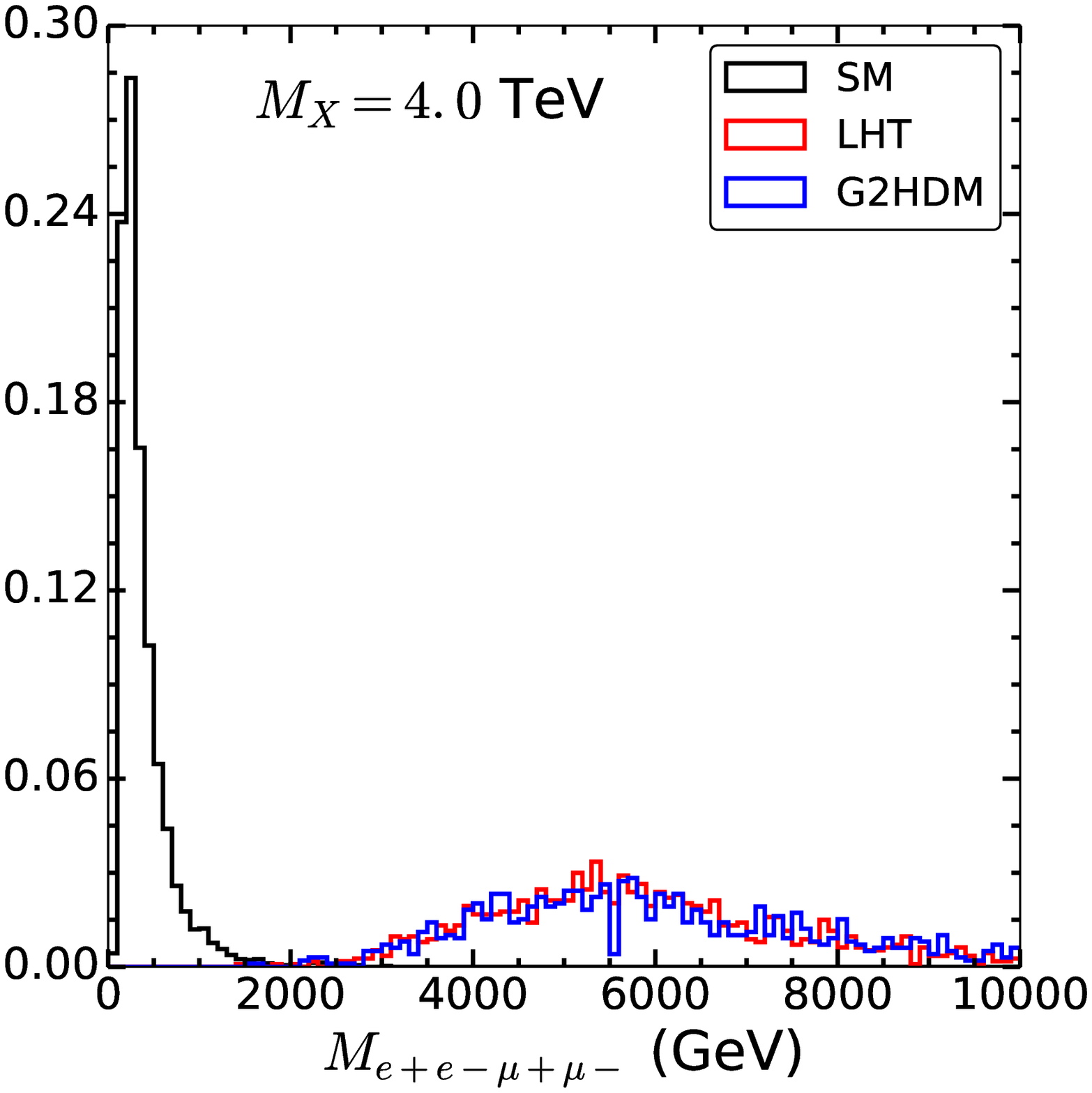}
\caption{\small \label{Fig:M4l}
Comparison of the normalized distributions of the invariant mass $ M_{e^{+}e^{-}\mu^{+}\mu^{-}} $ in
G2HDM (blue), LHT (red), and SM (black) computed at $ \sqrt{s}= $ 100 TeV 
with $ m_{X}=0.5\,\tev$ (top-left), $ 1.5\,\tev$ (top-right), $ 3.0\,\tev$ (bottom-left), and $ 4.0\,\tev$ (bottom-right).
}\end{figure}

Let $X$ denotes $W'^{(p,m)}, W^\pm_H$, or $Z_H$, notwithstanding 
the same symbol has been used for the $U(1)_X$ gauge boson 
which has been assumed to be very heavy and decoupled.
In four benchmark points $ m_{X}=0.5, \, 1.5, \, 3.0$, and $4.0\,\tev$, 
we will show the normalized kinematical distributions of the spatial separation 
$ \bigtriangleup R_{e^{+}e^{-}} $ (Fig.~\ref{Fig:dR}) 
and invariant mass $ M_{e^{+}e^{-}} $ (Fig.~\ref{Fig:IM}) of the electron pair in the 
$2l+\mET$ channel, and the invariant mass $ M_{e^{+}e^{-}\mu^{+}\mu^{-}} $ (Fig.~\ref{Fig:M4l}) 
of four leptons in $4l+\mET$ channel~\footnote{The spatial separation between particles is defined as 
$ \bigtriangleup R = \sqrt{(\bigtriangleup\eta)^2 +(\bigtriangleup\phi)^2} $, 
where $ \bigtriangleup\eta $ and $ \bigtriangleup\phi $ are the difference 
in pseudo-rapidity and the azimuthal angle, respectively.  
On the other hand, the invariant mass squared between particles is defined as 
$ M^2 = (\Sigma_i p_i)^2 $, where $ p_i $ is the four-momenta of particle $i$.}.
The muon has the same distributions of $ \bigtriangleup R_{\mu^{+}\mu^{-}} $ and
$ M_{\mu^{+}\mu^{-}}$ as the electron, and will not be discussed separately.
For a comparison, the benchmark point $m_X=0.5$ TeV 
is also included because its distribution shape is clearly distinguishable from the SM background.  
The corresponding coupling $g_H$ for $m_X=0.5\,\tev$ shall be appropriately small 
to avoid the current LHC limits as shown in Fig~\ref{fig:zp_con}.
The normalized kinematical distributions, nonetheless, do not depend on the values of $g_H$
that we shall keep in mind here and hereafter.
In addition, the distributions do not change significantly between the two spectra and
we simply choose Spectrum A.
The leading order irreducible SM background for each kinematical distribution is also 
presented for comparison. Further discussions of these three distributions are as follows.

First, the spatial separation $ \bigtriangleup R_{e^{+}e^{-}} $ distribution for $W'^{(p,m)}$ is distinct
from that of $W_{H}^{\pm}$.
As shown in Table~\ref{tab:signature}, the final lepton pair $l^+l^-$ is coming
from different decay patterns of the gauge bosons $W^{\prime p}W^{\prime m}$, $W^+_H W^-_H$, 
or $Z_H Z_H$ in the two models.
In G2HDM, both $l^+$ and $l^-$ are coming from either $W^{\prime p}$ or $W^{\prime m}$, 
while in LHT there are two possible routes -- 
(1) $l^+$ from $W^+_{H}$ and $l^-$ from 
$W_H^-$, or (2) both $l^+$ and $l^-$ from either one of the $Z_H$ in the $Z_H Z_H$ pair.
Overall, since $W^+_HW^-_H$ pair production has much larger cross section
than $Z_HZ_H$, for the two leptons 
in the final state, $l^+$ mainly comes from $W^+_H$ and $l^-$ from $W^-_H$, it leads to a larger $ \bigtriangleup R_{e^{+}e^{-}} $ in LHT than in G2HDM.  
Note that the distinction between the two models also depends on how 
boosted the gauge boson $X$ is.
If $X$ is highly boosted~({\it e.g.} $m_X=0.5$ TeV in Fig.~\ref{Fig:dR}), the distinction between
G2HDM and LHT becomes more visible.
For the SM case, the main contributions result from the $W^+W^-$ pair,
each of which decays into a charged lepton and a neutrino,
resulting in a large separation $ \bigtriangleup R_{e^{+}e^{-}} $ 
similar to the LHT case as is clearly reflected in Fig.~\ref{Fig:dR}.

Second, with the same reason as in the $ \bigtriangleup R_{e^{+}e^{-}} $ distribution,
the invariant mass of the lepton pair $ M_{e^{+}e^{-}} $ should be smaller than $ m_{W'^{(p,m)}} $ 
in G2HDM (or $ m_{Z_{H}} $ in LHT). 
From the position of the $ M_{e^{+}e^{-}} $ cut-off, 
one can roughly infer the mass $ m_{W'^{(p,m)}} $.
On the other hand, the invariant mass spectrum is distributed more 
evenly in the $ X = W^{\pm}_{H} $ case. 
For the SM, $M_{e^{+}e^{-}} $ is centered around $m_Z$ 
as well as the low mass region due to $ W^+W^- $ pair, off-shell $\gamma /Z$ 
and other non-resonance contributions.
The SM contribution is, of course, independent of $m_X$ 
and is shown only in the top-left panel of Fig.~\ref{Fig:IM} to highlight the difference 
from the new physics.
Therefore, one can simply impose appropriate cuts on $ M_{e^{+}e^{-}} $ to reduce the SM background 
and extract either G2HDM or LHT signals for different $ m_{X}$ at $ \sqrt{s}=100\,\tev$. 
From Fig.~\ref{Fig:IM}, one can clearly distinguish among the G2HDM, LHT, and SM
by the $ M_{e^{+}e^{-}} $ distributions. 

Last but not least, the cross section of $ 4l+\cancel{\it{E}}_{T} $ channel 
is about three times smaller than those of $ 2l +\cancel{\it{E}}_{T} $ for $W^{\prime \, (p,m)}$ 
searches in G2HDM as can be seen in the left panel of Fig.~\ref{Fig:Xsec}. 
Due to the facts that only $W^{\prime \, (p,m)}$ in G2HDM and 
$Z_H$ in LHT contribute to this channel and both of them are neutral, 
similar distributions of $ M_{e^{+}e^{-}\mu^{+}\mu^{-}} $ between G2HDM and LHT exhibited in Fig.~\ref{Fig:M4l} are expected.
However their distributions are clearly distinguishable from those of the SM which arise from on-shell $Z$ decays and consequently peak toward low-energy regions. Thus final state of four leptons plus missing transverse energy can be used to detect physics beyond the SM. 
However to distinguish G2HDM from LHT is not easy using the $ 4l+\cancel{\it{E}}_{T} $ channel 
unless $M_X$ is $\lesssim$ 0.5 TeV in which case a much smaller $g_H$ is anticipated.

From our studies of the two channels of two/four leptons plus missing transverse energy, one can conclude that 
in addition to the production cross sections, the kinematical distributions are also indispensable 
to discriminate the two models, G2HDM and LHT.

\section{Summary and Outlook}
\label{sec:summary}

One of the most interesting features in G2HDM~\cite{Huang:2015wts}, 
where the two Higgs doublets $H_1$ and $H_2$
are paired up to form a doublet of a new local $SU(2)_H$ gauge group,
is the three electrically neutral gauge bosons, $Z^\prime$ and $W'^{(p,m)}$. 
The $SU(2)_H$ is broken by the vevs of a triplet $\Delta_H$ and a doublet $\Phi_H$ 
from which a vev for the SM doublet $H_1$ is induced as well while 
$H_2$ is inert. 
As a result, all weak gauge bosons other than the photon got their masses.
The extra heavy fermions $Q^H$, $L^H$ and $\nu^H$, required by the anomaly cancellation
with the SM fermions, provide gauge invariant Yukawa couplings in Eqs.~(\ref{eq:YuK_new}) and (\ref{eq:Yuk_SM}) with the scalar doublets $H_1$, $H_2$, and $\Phi_H$ 
to give masses to all fermions. 
While $Z^\prime$ can decay into a pair of new heavy fermions, $W^{\prime (p,m)}$ can
only decay into a new heavy fermion and a SM fermion. 
The heavy fermion can decay into a SM fermion plus missing energy carried by the DM candidate
$H^0_2$, whose stability is protected by an emergent $Z_2$ symmetry.

In this work, we studied
collider signals of $Z^\prime$ and $W'^{(p,m)}$ which can help us to pinpoint G2HDM.
We derived constraints on $Z'$ from the LHC 13 TeV data,
followed by investigations of the future 14 TeV LH-LHC and 100 TeV proton-proton 
collider searches for $Z'$ and $W^{\prime (p,m)}$. 
The main difference between $Z^\prime$ and $W'^{(p,m)}$ is that $Z^\prime$ can be singly produced and decay into SM fermions, while $W^\prime$ has to be pair-produced in
light of the $SU(2)_H$ gauge symmetry. It leads to stringent bounds
on the $SU(2)_H$ coupling $g_H$ and the mass $m_{Z^\prime}$ from the LHC direct searches based 
on the Drell-Yan type dilepton and dijet final states.

The updated LHC limit shown on $(m_{Z'},g_H)$ plane in Fig.~\ref{fig:zp_con} 
is roughly a factor of two improvement in the low-mass
region $m_{Z^\prime} \lesssim 2.25$ TeV compared to the previous results~\cite{Huang:2015wts}
inferred from the LHC-8 dilepton data. In addition, for the whole region of $m_{Z'}$ of interest
the dilepton constraints on $g_H$ now becomes more stringent than those deduced from LEP and the electroweak
precision data. Even the dijet bounds which suffer from the QCD background are also stronger than the LEP bounds 
for $m_{Z'} \lesssim 2.5$ TeV.

Moreover, $Z'$ can also decay into the new fermions which subsequently decay into 
SM fermions plus the missing transverse energy.
We presented the contours of the rescaled cross sections $\sigma /g_H^2$
on the plane of the masses of $Z^\prime$ and new heavy exotic fermions
in Figs.~(\ref{fig:SPA}) and (\ref{fig:SPB}) for Spectrum-A$^\prime$ and Spectrum-B$^\prime$ respectively.
For Spectrum-A$^\prime$, we considered final states of two leptons~($e$s and $\mu$s) and two $\tau$s 
with missing transverse energy which originate from decays of the exotic heavy leptons. 
For Spectrum-B$^\prime$, two jets, two $b$-quarks and two $t$-quarks with missing transverse energy 
coming from the exotic heavy quarks were considered.
Using the LHC constraints on $\langle \epsilon \sigma \rangle^{95}_{\rm obs}$ from SUSY searches for the same
final states, one can derive limits for the coupling $g_H$ from the contours of $\sigma/g_H^2$ on the 
($m_{Z'}$, $m_{l^H}$) or ($m_{Z'}$, $m_{\tau^H}$) planes (Fig.~(\ref{fig:SPA})).
While the constraints on $g_H$ obtained from exotic decays of $Z'$ into heavy exotic leptons 
are comparable with those from the aforementioned dilepton and dijet searches, it appears that there is no severe 
bounds can be derived on the masses of heavy leptons from existing data of SUSY searches.
On the other hand, for Spectrum-B$^\prime$ in which the pair production of the exotic quarks  
is kinematically allowed has substantial regions being excluded (Fig.~(\ref{fig:SPB})). 
The reason is that the dominant contributions to the exotic quark production  arise from the QCD processes which are independent of the 
gauge coupling $g_H$. Thus, stringent limits from LHC on SUSY squark searches can be directly applied to our case, 
requiring the new quarks to be heavier than 1 TeV or so.      

\begin{table}[th!]
\begin{tabular}{|c|c|c|c|c|}
\hline
\multicolumn{2}{|c|}{Models} &Production &$2 l +\cancel{\it{E}}_{T}$ &$4 l +\cancel{\it{E}}_{T}$ \\
\hline
\multirow{2}*{G2HDM} &$Z'$ &S & Yes & No \\
\cline{2-5}
& $W'^{(p,m)}$ &P &Yes &Yes \\
\hline
\multirow{2}*{LHT} &$Z_H$ &P &Yes &Yes  \\
\cline{2-5}
&$W^\pm_H$ &P &Yes &No \\
\hline
\end{tabular}
\caption{Classification and search strategies of the gauge bosons in G2HDM and LHT
by single (S) / pair (P) productions and two decay channels.}\label{Tab:classification}
\end{table}

If $Z'$ can be discovered in the future, the neutral $W^{\prime (p,m)}$ also need to be found in order to 
identify G2HDM as the underlying theory. 
Final states of two and four leptons with missing transverse energy from G2HDM and LHT 
had been studied 
in detail to underline different signatures between the neutral $W^{\prime (p,m)}$ 
in G2HDM and the gauge bosons, $Z_H$ and $W^\pm_H$, in LHT.
The total cross sections for $ 2l +\cancel{\it{E}}_{T} $ and $ 4l +\cancel{\it{E}}_{T} $ 
were computed in both models~(Fig.~(\ref{Fig:Xsec})).
In the $ 2l +\cancel{\it{E}}_{T} $ channel, the two final leptons come from the 
same $W^{\prime (p,m)}$ in G2HDM, whereas in LHT they come from different
$W^\pm_H$.
Therefore, a smaller spatial separation between the final leptons is expected for 
$W^{\prime (p,m)}$. Indeed as clearly seen in Fig.~(\ref{Fig:dR}), 
the spatial separation of $e^+e^-$ 
in LHT is completely overlapped with the SM one which has larger $\Delta R_{e^+e^-}$.  
Furthermore, by the same reason, the invariant mass of the lepton pair are cut off at 
the mass of $W^{\prime (p,m)}$ as opposed to
a flatter invariant mass distribution for LHT~(Fig.~(\ref{Fig:IM})).
In addition, the $ 4l +\cancel{\it{E}}_{T} $ channel is also investigated.
The invariant mass distributions of four charged leptons behave quite differently 
between the SM and G2HDM~(and LHT), whereas LHT and G2HDM exhibit similar distributions~(Fig.~(\ref{Fig:M4l})). 

We conclude our study by presenting the search strategies 
of distinguishing G2HDM from LHT in Table~\ref{Tab:classification}. 
For $Z^\prime$ which can be singly produced, the dilepton final states will be the best search channels.  
For pair-produced gauge bosons like $W^{\prime (p,m)}$ in G2HDM, and $W^\pm_H$ and $Z_H$ in LHT, 
apart from the total cross sections, detailed kinematical distributions,
such as $(i)$ the spatial separation between the SM lepton pair
and $(ii)$ the invariant mass distributions of two and four leptons 
in the final states, can help us to disentangle these two models. 

High luminosity upgrade for the LHC and building a future 100 TeV hadron collider are 
matters of utmost importance for fully exploring and distinguishing new electroweak scale 
models, all of which are contrived to address theoretical issues that cannot be answered within the SM.

\newpage

\appendix
\section{Scalar Potential of G2HDM}\label{section:app} 

We here spell out the most general and renormalizable scalar potential invariant under the $SU(2)_H \times U(1)_X$ as well as the SM gauge groups,
given the particle contents specified in Table~\ref{tab:quantumnos}. 
There exist two Higgs doublets, $H_1$ and $H_2$ where $H_1$ is identified as the SM Higgs doublet and $H_2$~(with the same hypercharge $Y=1/2$ as $H_1$)
is the additional $SU(2)_L$ scalar doublet.
The two Higgs doublets $H_1$ and $H_2$ are embedded into an $SU(2)_H$ doublet $H=(H_1\; H_2)^T$
with a $U(1)_X$ charge of $X(H)=1$. 

In addition to $H$, we introduce the $SU(2)_H$ triplet $\De_H$ and doublet $\Phi_H$, 
 \begin{align}
&\De_H =
  \begin{pmatrix}
    \De_3/2   &  \De_p / \sqrt{2}  \\
    \De_m / \sqrt{2} & - \De_3/2   \\
  \end{pmatrix} \; = \De^\dagger_H \;\; , \;\; 
  \Phi_H =
  \begin{pmatrix}
      \Phi_1  \\
     \Phi_2   \\
  \end{pmatrix} \; , 
 \end{align} 
with $\Delta_m = \left( \Delta_p \right)^* \; {\rm and} \; \left( \Delta_3 \right)^* = \Delta_3$. Both of them are {\it singlets} under the SM gauge groups. 
The vev of the triplet can induce $SU(2)_L$ symmetry breaking as we shall see below, while that of $\Phi_H$ provides a mass to the new
fermions that are necessitated due to the $SU(2)_H$-invariant Yukawa couplings
shown in Eqs.~\eqref{eq:YuK_new} and \eqref{eq:Yuk_SM}.
In the following, the $SU(2)_L$ multiplication is implicitly assumed but not denoted unless otherwise stated.

The scalar potential invariant under
$SU(2)_L\times U(1)_Y \times SU(2)_H \times U(1)_X$ reads
\begin{align}
V \left( H , \Delta_H, \Phi_H \right) = V (H) + V (\Phi_H ) + V ( \De_H ) + V_{\rm mix} \left( H , \Delta_H, \Phi_H \right) \; , 
\label{eq:higgs_pot} 
\end{align}
with
\begin{align}
\label{VH1H2}
V (H) 
=& \; \mu^2_H   \left( H^{\dag\alpha i}  H_{\alpha i} \right)
+  \la_H \left( H^{\dag\alpha i}  H_{\alpha i} \right)^2  
+ \frac{1}{2} \la'_H \epsilon_{\alpha \beta} \epsilon^{\gamma \delta}
\left( H^{\dag \alpha i}  H_{\gamma  i} \right)  \left( H^{\dag \beta j}  H_{\delta j} \right)  \; , \nn \\
=&  \;  \mu^2_H   \left( H^\dag_1 H_1 + H^\dag_2 H_2 \right) 
+ \la_H   \left( H^\dag_1 H_1 + H^\dag_2 H_2 \right)^2 
+ \la'_H \left( - H^\dag_1 H_1 H^\dag_2 H_2 
                  + H^\dag_1 H_2 H^\dag_2 H_1 \right)  \; , 
\end{align}
where $\alpha$, $\beta$, $\gamma$, and $\delta$~($i,j$) refer to the $SU(2)_H$~($SU(2)_L$) indices;
all of them run from one to two and the superscript~(subscript) is always attached to $H^\dag ~(H)$.
In light of the equality,
$\epsilon_{\alpha\beta}\epsilon^{\gamma\delta}=
- \delta_{\alpha}^{\gamma} \delta_{\beta}^{\delta}
+ \delta_{\alpha}^{\delta} \delta_{\beta}^{\gamma}$,
one can express $\epsilon_{\alpha \beta} \epsilon^{\gamma \delta} 
\left( H^{\dag \alpha i}  H_{\gamma  i} \right)  \left( H^{\dag \beta j}  H_{\delta j} \right)$
as a linear combination of two independent terms: 
\begin{align}
\left( H^{\dag\alpha i}  H_{\alpha i} \right)  \left( H^{\dag\beta j}  H_{\beta j} \right) \;\; \text{and} \;\;
\left( H^{\dag\alpha i}  H_{\beta i} \right) \left( H^{\dag\beta j}  H_{\alpha j} \right) .
\end{align}
An easy way to see that these two  terms are the only possibilities is to notice that for quartic terms in $H$ and $H^\dag$ one always requires two of $H^\dag$s and two of $H$s to obey the $U(1)_Y$ and $U(1)_X$ symmetry.
In this case, one is left with only two options in terms of $H$: $H_{\alpha i} H_{\beta j}$ and $H_{\alpha j} H_{\beta i}$.
The first option yields $\left( H^{\dag\alpha i}  H_{\alpha i} \right)  \left( H^{\dag\beta j}  H_{\beta j} \right) $ after gauge index contractions,
while the second one is $\left( H^{\dag\alpha i}  H_{\beta i} \right) \left( H^{\dag\beta j}  H_{\alpha j} \right) $.
Any other different $SU(2)$ combinations with either antisymmetric $\epsilon$~(Levi-Civita symbol) or
$\delta$~(Kronecker delta) tensor can be rewritten as functions
 of these two terms. As for $V ( \Phi_H )$ and $V ( \Phi_H )$, one has
\begin{align}
\label{VPhi}
V ( \Phi_H ) =& \;  \mu^2_{\Phi}   \Phi_H^\dag \Phi_H  + \la_\Phi \left( \Phi_H^\dag \Phi_H  \right)^2  \;  \nn \\
 =& \;  \mu^2_{\Phi} \left( \Phi^*_1\Phi_1 + \Phi^*_2\Phi_2 \right) 
 +  \la_\Phi \left( \Phi^*_1\Phi_1 + \Phi^*_2\Phi_2 \right)^2 \; , \\
 \label{VDelta}
V ( \De_H ) =& \; - \mu^2_{\De} {\rm Tr} \left( \De^2_H  \right) 
 \;  + \la_\De \left( {\rm Tr} \left( \De^2_H  \right) \right)^2 
 \;  \nn \\
= & \; - \mu^2_{\De} \left( \frac{1}{2} \De^2_3 + \De_p \De_m  \right) +  \la_{\De} \left( \frac{1}{2} \De^2_3 + \De_p \De_m  \right)^2 \; . 
\end{align}
The trace of  terms with odd powers in $\De_H$ is vanishing.
In addition, there exists another quartic term in $\De_H$, ${\rm Tr} \left( \De^4_H\right)$ that, however, is not independent as $\left( {\rm Tr} \left( \De^2_H  \right) \right)^2  = 2 \, {\rm Tr} \left( \De^4_H\right)$. 
Finally, the potential with mixed terms is
\begin{align}
\label{VMix}
V_{\rm{mix}} \left( H , \Delta_H, \Phi_H \right) = 
& \; + M_{H\De}  \left( H^\dag \De_H H \right) -  M_{\Phi\De}  \left( \Phi_H^\dag \De_H \Phi_H \right)  \nn \\
& \; + \la_{H\Phi} \left( H^\dag H  \right)  \left( \Phi_H^\dag \Phi_H \right)  
 + \la^\prime_{H\Phi} \left( H^\dag \Phi_H  \right)  \left( \Phi_H^\dag H \right)
\nn\\
& \;  +  \la_{H\De} \left( H^\dag H  \right)    {\rm Tr} \left( \De^2_H  \right)  
+ \la_{\Phi\De} \left( \Phi_H^\dag \Phi_H \right) {\rm Tr} \left( \De^2_H \right)  \;  \nn \\
= & \; + M_{H\De} \left( \frac{1}{\sqrt{2}}H^\dag_1 H_2 \De_p  
+  \frac{1}{2} H^\dag_1 H_1\De_3 + \frac{1}{\sqrt{2}}  H^\dag_2 H_1 \De_m  
- \frac{1}{2} H^\dag_2 H_2 \De_3   \right)   \nn \\
& \; - M_{\Phi\De} \left(  \frac{1}{\sqrt{2}} \Phi^*_1 \Phi_2 \De_p  
+  \frac{1}{2} \Phi^*_1 \Phi_1\De_3 + \frac{1}{\sqrt{2}} \Phi^*_2 \Phi_1 \De_m  
- \frac{1}{2} \Phi^*_2 \Phi_2 \De_3   \right)  \nn \\
& \; +  \la_{H\Phi} \left( H^\dag_1 H_1 + H^\dag_2 H_2 \right)  \left( \Phi^*_1\Phi_1 + \Phi^*_2\Phi_2 \right) \nn\\
& \; +  \la^\prime_{H\Phi} \left( H^\dag_1 H_1 \Phi^*_1\Phi_1 + H^\dag_2 H_2  \Phi^*_2\Phi_2 
+ H^\dag_1 H_2 \Phi_2^*\Phi_1 + H^\dag_2 H_1  \Phi^*_1\Phi_2  \right) \nn\\
& \; + \la_{H\De} \left( H^\dag_1 H_1 + H^\dag_2 H_2 \right)   \left( \frac{1}{2} \De^2_3 + \De_p \De_m  \right) \nn\\
& \; + \la_{\Phi\De}  
  \left(  \Phi^*_1\Phi_1 + \Phi^*_2\Phi_2 \right)  \left( \frac{1}{2} \De^2_3 + \De_p \De_m  \right)  \; .
\end{align}
Note that extra terms $ \left( H^\dag \De^2_H  H  \right)$ and $\left( \Phi_H^\dag \De^2_H \Phi_H \right) $ are not independent but instead proportional to $\left( H^\dag H  \right)    {\rm Tr} \left( \De^2_H  \right)$ and $\left( \Phi_H^\dag \Phi_H \right) {\rm Tr} \left( \De^2_H \right)$ respectively, while $\left( \Phi_H^T  \epsilon H \right)^\dag \left( \Phi_H^T \epsilon H \right)$ can be expressed as $\left( H^\dag H  \right)  \left( \Phi_H^\dag \Phi_H \right)  - \left( H^\dag \Phi_H  \right)  \left( \Phi_H^\dag H \right)$.

We should point out that the $\la'_H$ in Eq.~\eqref{VH1H2}
the $\lambda^{\prime}_{H\Phi}$ term in Eq.~\eqref{VMix} were not included in 
the original work~\cite{Huang:2015wts}.
Moreover, all dimensionful mass parameters of the cubic couplings 
and dimensionless quartic couplings are necessarily {\it real} because every term in the mixed potential
$V_{\rm mix}(H, \Delta_H,\Phi_H)$ in (\ref{VMix}) is Hermitian.

The resulting coefficients of the quadratic terms for $H_1$ and $H_2$ after spontaneous
symmetry breaking of $SU(2)_H$ induced by
 $\langle \Delta_3 \rangle = - v_\Delta$ are
\begin{align}
\label{eq:H1vev}
&\mu^2_H - \frac{1}{2} M_{H\Delta} \cdot v_\Delta + \frac{1}{2} \lambda_{H \Delta} \cdot v_\Delta^2 
+  \cdots \; , \\
\label{eq:H2vev}
&\mu^2_H + \frac{1}{2} M_{H\Delta} \cdot v_\Delta + \frac{1}{2} \lambda_{H \Delta} \cdot v_\Delta^2 
+  \cdots \; , 
\end{align} 
where ``$\cdots$'' refers to terms not containing $v_\Delta$.
As a consequence, even with a positive $\mu_H^2$, the breaking of $SU(2)_H$ with $v_\Delta \neq 0$ 
can trigger the breaking of $SU(2)_L$ to give rise $\langle H_1 \rangle \neq 0$ if the sum of the second 
and third terms in~(\ref{eq:H1vev}) can be sufficiently negative~\cite{Arhrib:2018sbz}.
On the other hand, $H_2$ that does not develop a vev can play a role of the inert Higgs doublet.   

\section{Scalar Mass Spectra}\label{section:app_spectrum} 

In light of spontaneous symmetry breaking, we reparametrize the fields as 
\begin{eqnarray}
H_1 = 
\begin{pmatrix}
G^+ \\ \frac{v + h}{\sqrt 2} + i \frac{G^0}{\sqrt 2}
\end{pmatrix}
, \;
H_2 = 
\begin{pmatrix}
H^+ \\ H_2^0
\end{pmatrix}
, \;
\Phi_H = 
\begin{pmatrix}
G_H^p \\ \frac{v_\Phi + \phi_2}{\sqrt 2} + i \frac{G_H^0}{\sqrt 2}
\end{pmatrix}
, \;
\Delta_H =
\begin{pmatrix}
\frac{-v_\De + \delta_3}{2} & \frac{1}{\sqrt 2}\De_p \\ 
\frac{1}{\sqrt 2}\De_m & \frac{v_\De - \delta_3}{2}
\end{pmatrix}\nn \\
.
\end{eqnarray}
The scalar boson mass can be attained by taking the second derivatives of 
the potential in Eq.~\eqref{eq:higgs_pot} with respect to the corresponding scalar field, evaluated around the vacuum.
There are mixing terms among the fields, depending on their quantum numbers.
We start with a 3-by-3 mass matrix, consisting of three real scalars $S=\{h, \phi_2, \delta_3\}$ 
\begin{align}
{\mathcal M}_0^2 =
\begin{pmatrix}
2 \la_H v^2 & \la_{H\Phi} v v_\Phi 
& \frac{v}{2} \left( M_{H\De} - 2 \la_{H \De} v_\De \right)  \\
\la_{H\Phi} v v_\Phi
& 
2 \la_\Phi v_\Phi^2
&  \frac{ v_\Phi}{2} \left( M_{\Phi\De} - 2 \la_{\Phi \De} v_\De \right) \\
\frac{v}{2} \left( M_{H\De} - 2 \la_{H \De} v_\De \right)  & \frac{ v_\Phi}{2} \left( M_{\Phi\De} - 2 \la_{\Phi \De} v_\De \right) & \frac{1}{4 v_\De} \left( 8 \la_\De v_\De^3 + M_{H\Delta} v^2 + M_{\Phi \De} v_\Phi^2 \right)   
\end{pmatrix} \; .
\label{eq:scalarbosonmassmatrix}
\end{align}
This matrix can be diagonalized by an orthogonal matrix  $O$, defined as
$ \vert f \rangle_i \equiv O_{i j}  \vert m \rangle_j $ with $i$ and $j$ representing the flavor and mass eigenstates respectively,  
\begin{equation}
O^T \cdot {\mathcal M}_0^2 \cdot O = {\rm Diag}(m^2_{h_1}, m^2_{h_2}, m^2_{h_3}) \; ,
\end{equation}
where the three eigenvalues are in ascending order.
Since we focus on the situation of $v_\Delta \sim v_\Phi \gg v$,
the lightest eigenstate with a mass $m_{h_1}$ will be identified as the 125 GeV Higgs boson discovered at the LHC 
and the other two heavier Higgses $h_2$ and $h_3$ have the mass of $m_{h_2}$ and $m_{h_3}$. 
The observed 125 GeV Higgs boson is a mixture of the three neutral components 
$h$, $\phi_2$ and $\delta_3$.

The second mass matrix, comprised of three complex scalars
$G=\{ G^p_H , H^{0*}_2, \De_p  \}$, is 
\begin{align}
{\mathcal M}_0^{\prime 2} =
\begin{pmatrix}
M_{\Phi \De} v_\De  +\frac{1}{2}\la^\prime_{H\Phi}v^2 & \frac{1}{2}\la^\prime_{H\Phi}  v v_\Phi & - \frac{1}{2} M_{\Phi \De} v_\Phi  \\
\frac{1}{2}\la^\prime_{H\Phi} v v_\Phi &  M_{H \De} v_\De 
+\frac{1}{2}\la^\prime_{H\Phi} v_\Phi^2
 &  
\frac{1}{2} M_{H \De} v\\
- \frac{1}{2} M_{\Phi \De} v_\Phi & \frac{1}{2} M_{H \De} v & 
\frac{1}{4 v_\De} \left( M_{H\De} v^2 + M_{\Phi \De} v_\Phi^2 \right)\end{pmatrix} \; .
\label{goldstonemassmatrix}
\end{align}
This matrix has a zero eigenvalue as can be seen by the vanishing determinant and that eigenstate corresponds to the 
physical Goldstone bosons,  $\widetilde G^{p,m} \sim v_\Phi G^{p,m}_H - v H^{0*,0}_2 + 2 v_\De \De_{p,m}$.
The other two eigenvalues are the masses of two physical complex fields $\widetilde \Delta$ and $D$ as
\begin{eqnarray}
\label{darkmattermass}
M^2_{{\widetilde \Delta},D} &=& \frac{-B \pm \sqrt{B^2 - 4 A C}}{2A} \; ,
\end{eqnarray}
with 
\begin{eqnarray}
\label{ABC}
A & =& 8 \, v_\Delta \; , \nonumber \\
B & = & - 2 \left( M_{H\Delta} \left( v^2 + 4 v_\Delta^2 \right) + M_{\Phi \Delta} \left( 4 v_\Delta^2 + v_\Phi^2 \right)
+ 2 \la^\prime_{H\Phi} v_\Delta \left( v^2 + v_\Phi^2 \right) \right) \;, \\
C & = & \left( v^2 + v_\Phi^2 + 4 v_\Delta^2 \right) 
\left( M_{H \Delta} \left( \la^\prime_{H\Phi} v^2 + 
2 M_{\Phi \Delta} v_\Delta \right) + \la^\prime_{H\Phi} M_{\Phi \Delta} v_\Phi^2  \right) \; .\nonumber
\end{eqnarray}
The mass eigenstate $D$ can be a DM candidate in G2HDM, accounting for the correct relic density~\cite{Huang:2015wts}. 
Besides, there are four Goldstone boson fields $G^\pm$, $G^0$ and $G^0_H$
with masses 
\begin{equation}
m^2_{G^\pm} = m^2_{G^0} = m^2_{G^0_H} = 0 \; ,
\end{equation}
and a physical charged Higgs boson $H^\pm$ with a mass
\begin{equation}
m^2_{H^\pm} = M_{H \De} v_\De  - \frac{1}{2}\la^\prime_H v^2 +\frac{1}{2}\la^\prime_{H\Phi}v_\Phi^2 \; ,
\label{chargedHiggsmass}
\end{equation}

The six Goldstone particles, $G^\pm$, $G^0$, $G^0_H$ and $\widetilde G^{p,m}$, will be absorbed 
by the longitudinal components of the massive gauge bosons 
$W^\pm$, $Z$, $Z^\prime$ and $W^{\prime (p,m)}$, respectively.
As a result, one has two unbroken generators and two massless gauge bosons
left over after spontaneous symmetry breaking. 
One of them is naturally identified as the photon and the other one could play a role of either the
light dark photon $\gamma_D$ or heavy $Z^{\prime\prime}$.
To render the dark photon $\gamma_D$ or $Z^{\prime\prime}$ massive, one can either resort to the Stueckelberg 
mechanism or introduce yet another Higgs field $\Phi_X$ solely charged under $U(1)_X$ to 
break one of the remaining two unbroken generators.

\begin{table} 
\scriptsize
\centering
\begin{tabular}{|l|l|l|l|l|}
\hline
\hline
& \multicolumn{2}{|c|} {Spectrum-A (B)} &  \multicolumn{2}{|c|} {Spectrum-A$^\prime$ (B$^\prime$)} \\
\hline
\multicolumn{5}{|c|}{Input Parameters}\\
\hline

$g_{H}$ & 0.08 & 0.08 & 0.04 & 0.04 \\
$\lambda_H $ & 0.149 & 0.786 & 0.286 & 0.172 \\
$\lambda_\Phi $ & 1.455 & 0.237 & 0.546 & 0.908 \\
$\lambda_\Delta$ & 4.557 & 2.840 & 0.786 & 0.555 \\
$\lambda_{\Phi\Delta}$ & 0.616 & 0.346 & 0.198 & 0.199 \\
$\lambda_{H\Delta}$ & 0.360 & 0.570 & $-0.150$ & $-0.237$ \\
$\lambda_{H\Phi}$ & 0.297 & 0.789 & 0.542 & 0.199 \\
$\lambda_{H\Phi}^\prime$ & 0.383 & 0.992 & 0.875 & 0.466 \\
$\lambda_{H}^\prime$ & $-0.048$ & 0.455 & $-0.005$ & -0.346 \\
$M_{\Phi\Delta}$ (\gev) & 0.297 & 0.365 & 0.003 & 0.005 \\
$M_{H\Delta}$ (\gev) & 73.954 & 11.200 & 45.799 & 23.313 \\
$v$ (\gev) & 246.000 & 246.000 & 246.000 & 246.000 \\
$v_{\Phi}$ (\tev) & 67.329 & 48.521 & 88.157 & 77.607 \\
$v_{\Delta}$ (\tev) & 1.074 & 1.329 & 3.076 & 3.485 \\

\hline
\multicolumn{5}{|c|}{Mass Spectrum}\\
\hline

$m_{h_1}$ (\gev) & 125.489 & 125.105 & 125.526 & 125.188 \\
$m_{h_2}$ (\tev) & 3.268 & 3.122 & 3.813 & 3.636 \\
$m_{h_3}$ (\tev) & 114.855 & 33.414 & 92.125 & 104.584 \\
$m_{H^\pm}$ (\tev) & 29.465 & 34.172 & 58.312 & 37.462\\
$M_{D}$ (\gev) & 561.035 & 402.815 & 46.157 & 47.746 \\
$M_{\tilde\Delta}$ (\tev) & 29.465 & 34.173 & 58.312 & 37.462 \\
$m_{Z'}$ (\tev) & 2.693 & 1.941 & 1.763 & 1.552 \\
$m_{W^{\prime (p,m)}}$ (\tev) & 2.695 & 1.944 & 1.767 & 1.558 \\

\hline
\hline
\end{tabular}
\caption{\label{tab:inputs} 
Four benchmark points in the parameter space for the four mass spectra 
of interest in this work. The gauge coupling $g_H$ is fixed to be either $0.08$ or $0.04$  
in order to satisfy the Drell-Yan constraints, shown in Fig.~\ref{fig:zp_con}.}
\end{table}
 

In this work, we confine ourselves to the scenarios with $ m_{Z'}\simeq m_{W^{\prime (p,m)}} \gtrsim 5 \,m_D$
and the heavy fermion masses of $\mathcal{O(\text{TeV})}$, which are determined by the new Yukawa couplings
and $\langle \Phi_2 \rangle$: $m_{f^H}=y^\prime_f v_\Phi /\sqrt{2}$~($f=d,u,e,\nu$) from Eq.~\eqref{eq:YuK_new}.
In order to realize the region of interest $1.5 \leq m_{Z'}\simeq m_{W^{\prime (p,m)}} \leq 3 $ TeV and avoid the LHC
$Z'$ bounds shown in Fig.~\ref{fig:zp_con}, from Eq.~\eqref{eq:zw_mas} the vevs of the $SU(2)_H$ doublet and triplet
have to be
\begin{align}
v_\Phi \gtrsim 40 \, \text{TeV} \gtrsim 3 v_\Delta  \, ,
\label{eq:v_phi_delta}
\end{align}
implies $y^\prime_f \lesssim 0.1$. 
In Table~\ref{tab:inputs}, we present four different benchmark points: 
the second and third columns are for either Spectrum-A or Spectrum-B, while the last two
columns are for either Spectrum-A$^\prime$ or Spectrum-B$^\prime$. 
Recall that, on one hand, the Spectrum-A and A$^\prime$ are with lighter new quarks while
the Spectrum-B and B$^\prime$ are with heavier new quarks.
The required Yukawa couplings $y^\prime_f$ for new fermions in G2HDM  
are of order $10^{-3}$ for Spectrum-A (A$^\prime$) 
and order $10^{-2}$ for Spectrum-B (B$^\prime$) which are quite acceptable.
On the other hand, the mass spectrum of the Higgses is the same for Spectrum-A and Spectrum-B 
and for Spectrum-A$^\prime$ and Spectrum-B$^\prime$.   
Similarly cases for the mass spectrum of new gauge bosons.
We should point out that the mass spectrum displayed here are composed of points in the parameter space 
that satisfy the two theoretical constraints -- 
the scalar potential is bounded from below and all relevant $2 \to 2$ scattering amplitudes among 
the scalars are below the unitarity bound. 
In addition to the theoretical constraints, we also impose the experimental constraints from the 
125 GeV Higgs, including its mass and signal strengths decaying into diphoton and $\tau^+\tau^-$.
Detailed study of these constraints on the scalar sector of G2HDM 
is presented in a separate work~\cite{Arhrib:2018sbz}. 
One should be aware that we choose a parameter space slightly different from that
presented in Ref.~\cite{Arhrib:2018sbz} where the new gauge sector is much heavier than 
the Higgs sector. In this analysis, we choose the new gauge bosons $Z^\prime$ and $W^{\prime (p,m)}$ 
having masses between $1.5\tev$ to $3\tev$ such that they are accessible at the LHC.
Nevertheless the parameter space in both works is chosen to satisfy the same set of aforementioned constraints.


\newpage

\noindent \section*{Acknowledgement}
We thank Van Que Tran and Raymundo Ramos for sharing with us the scan data and useful discussions.
TCY would like to thank the hospitality he received at NCTS where ideas of this project 
were nurtured.
CTL would like to thank Dr. Jung Chang for some useful suggestions for the simulation of littlest Higgs models
at the beginning of this project. We would like to thank the referee for pointing out a missing term in the scalar potential.
The work of TCY is supported by the Ministry of Science and Technology (MoST) of Taiwan under
Grant Number MOST-104-2112-M-001-001-MY3. 
WCH is supported by DGF Grant No. PA 803/10-1 and by Danish Council for Independent Research
Grant DFF-6108-00623.


\end{document}